\documentclass{nature}
\usepackage{notation}

\usepackage{graphicx}
\usepackage{tabularx}
\usepackage{CJKutf8}
\usepackage[mathscr]{euscript}
\usepackage{bm}
\usepackage{orcidlink}
\usepackage{aas_macros}
\usepackage{makecell}
\usepackage{svg}

\graphicspath{{figs/}}

\usepackage[acronym,numberedsection]{glossaries}

\newignoredglossary{ignored}
\makeglossaries

\newglossaryentry{sbc}
{
    name=Simulation-Based Calibration,
    description={Techniques that use simulations from the model to check the correctness of Bayesian computation\cite{doi:10.1080/01621459.1982.10477856, doi:10.1198/106186006X136976, 2018arXiv180406788T}}
}

\newglossaryentry{cod}
{
    name=curse of dimensionality,
    description={The phenomenon that the difficulty of a problem often increases dramatically with dimension\cite{bellman1961adaptive}}
}

\newglossaryentry{Markov chain}
{
    name=Markov chain,
    description={A sequence of random states for which the probability of a state depends only on the previous state}
}

\newglossaryentry{integrand}
{
    name=integrand,
    description={A function that is being integrated, for example, $f(x)$ is the integrand in $\int f(x) \diff x$}
}

\newglossaryentry{prior}
{
    name=prior,
    description={A probability density that isn't yet conditioned on observed data. In NS it is a measure in our integral}
}

\newglossaryentry{constrained prior}
{
    name=constrained prior,
    description={The \gls{prior} for the parameters restricted to the region in which the \gls{likelihood} exceeds a threshold}
}

\newglossaryentry{posterior}
{
    name=posterior,
    description={A probability density conditioned on the observed data found by updating a \gls{prior} using Bayes' theorem. In NS it describes the shape of our \gls{integrand}}
}

\newglossaryentry{likelihood}
{
    name=likelihood,
    description={The probability of the observed data as a function of a model's parameters}
}

\newglossaryentry{Bayesian model comparison}
{
    name=Bayesian model comparison,
    description={A method for comparing models based on computing the change in their relative plausibility in light of data using Bayes' theorem}
}

\newglossaryentry{evidence}
{
    name=evidence,
    description={A factor in Bayes' theorem that may be written as an integral and that plays an important role in \gls{Bayesian model comparison}. In NS this is the multi-dimensional integral we wish to compute}
}

\newglossaryentry{parameter domain}
{
    name=parameter domain,
    description={The set of a priori possible parameters, usually the reals $\mathbb{R}^n$ or a subset thereof}
}

\newglossaryentry{survival function}
{
    name=survival function,
    description={A function $F(x)$ associated with a distribution that returns the probability of obtaining a sample greater than $x$}
}

\newglossaryentry{overloaded}
{
    name=overloaded,
    description={A function with a definition that depends on the type of its argument. In NS the likelihood $\like$ is an overloaded function since we consider separate functions $\like(\theta)$ and $\like(X)$}
}

\newglossaryentry{measure}
{
    name=measure,
    description={A probability measure is a function assigning probabilities to events. See \ccite{applebaum2008probability} for a pedagogical discussion}
}

\newglossaryentry{super-level-sets}
{
    name=super-level-sets,
    description={A $\lambda$-super-level-set of any function contains all points for which the function value exceeds $\lambda$}
}

\newglossaryentry{push-forward measure}
{
    name=push-forward measure,
    description={The push-forward measure (or image measure) is the distribution of a random variable under a probability measure
}
}

\newglossaryentry{geometrical triangulation}
{
    name=geometrical triangulation,
    description={The division of an object into triangles or simplexes}
}

\newglossaryentry{partition function}
{
    name=partition function,
    description={Normalizing constant $\ev(\beta) = \int e^{-\beta E(\params)} \diff \params$ that fully characterizes a physical system, because many important thermodynamic variables can be derived from it}
}

\newglossaryentry{law of large numbers}
{
    name=law of large numbers,
    description={Laws showing that the average of a quantity tends towards its true value as the number of samples increases},
    type={ignored}
}

\newglossaryentry{Kullback-Leibler divergence}
{
    name=Kullback-Leibler divergence,
    text={KL divergence},
    first={Kullback-Leibler (KL) divergence},
    description={A measure of difference between two distributions that may be interpreted as the information gained by switching from one to the other. See \ccite{applebaum2008probability,mackay2003information} for pedagogical discussions}
}

\newglossaryentry{iso-likelihood contour}
{
    name=iso-likelihood contour,
    description={The set of points for which the likelihood is equal to a particular constant; in two-dimensions, this set forms a contour line}
}

\newglossaryentry{bootstrap}
{
    name=bootstrap,
    description={Techniques that estimate statistical variation by repeatedly drawing samples from the true dataset with replacement}
}

\newglossaryentry{microcanonical ensemble}
{
    name=microcanonical ensemble,
    description={Assigns equal probability to states $\params$ with $E(\params) = \epsilon$ and zero probability if $E(\params) \not= \epsilon$ such that the energy level $\epsilon$ rather than the  inverse temperature $\beta$ characterizes a thermodynamic state}
}

\newglossaryentry{microstate}
{
    name=microstate,
    description={The state of all degrees of freedom in a physical system, for example, the microstate of a multi-particle system includes the positions and momenta of all particles}
}

\newglossaryentry{indicator function}
{
    name=indicator function,
    description={A function $\mathbb{1}(\cdot)$ that takes the value 1 if the condition $\cdot$ holds and 0 otherwise}
}

\newglossaryentry{mode}
{
    name=mode,
    description={A peak in a probability distribution}
}

\newglossaryentry{multi-modal}
{
    name=multi-modal,
    description={Problems in which the \gls{integrand} contains more than one \gls{mode}. In inference problems, \glspl{mode} may correspond to distinct ways in which the model could predict the observed data}
}

\newglossaryentry{uni-modal}
{
    name=uni-modal,
    symbol=uni-,  
    description={Problems in which the \gls{integrand} contains only one \gls{mode} (cf.~\gls{multi-modal})}
}

\newglossaryentry{transition kernel}
{
    name=transition kernel,
    description={A function that describes the likely steps of a \gls{Markov chain}}
}

\newglossaryentry{insertion indexes}
{
    name=insertion indexes,
    description={The indexes at which the elements of a list must be inserted into an ordered list to maintain ordering}
}

\newglossaryentry{Markov chain Monte Carlo}
{
  text={MCMC},
  description={A class of algorithms for drawing a correlated sequence of samples from a distribution using the states of a \gls{Markov chain}},
  name={Markov chain Monte Carlo (MCMC)},
  first={Markov chain Monte Carlo (MCMC; see for example \ccite{brooks2011handbook,Hogg:2017akh})}
}

\newglossaryentry{SZ}
{
    type={ignored}, 
    text=SZ,
    description={TODO},
    name={Sunyaev–Zel’dovich effect (SZ)},
    first={Sunyaev–Zel’dovich effect (SZ)}
}

\newglossaryentry{standard candles}
{
    type={ignored}, 
    name=standard candles,
    description={TODO}
}

\newglossaryentry{tractable}
{
    name=tractable,
    description={An adjective used to describe problems that are feasible to solve under computational, monetary or time constraints}
}

\newglossaryentry{pseudo-importance sampling}
{
    name=pseudo-importance sampling,
    description={Algorithms in which an importance sampling density is defined a posteriori}
}

\newglossaryentry{bulk}
{
    name=bulk,
    description={Region with size of order $e^{-\kl}$ that contains the overwhelming majority of the posterior mass. Closely related to typical sets. Usually won't lie near the mode of the posterior, especially in high-dimensions~\cite{mackay2003information,betancourt2018conceptual}}
}

\begin{document}
\newcommand{\separator}{{${}^{,}$}}

\title{Nested Sampling for physical scientists}
\author{Greg Ashton\,\orcidlink{0000-0001-7288-2231},\authornumber{1}\separator\authornumber{2}
Noam Bernstein\,\orcidlink{0000-0002-6532-1337},\authornumber{3}
Johannes Buchner\,\orcidlink{0000-0003-0426-6634},\authornumber{4}
Xi Chen\,\orcidlink{0000-0002-3577-3308},\authornumber{5}
G\'{a}bor Cs\'{a}nyi\,\orcidlink{0000-0002-8180-2034},\authornumber{6}
Farhan Feroz\,\orcidlink{0000-0001-6726-0032},\authornumber{7}
Andrew Fowlie\,\orcidlink{0000-0001-5457-6329},\authornumber{8}
Matthew Griffiths\,\orcidlink{0000-0002-2553-2447},\authornumber{9}
Michael Habeck\,\orcidlink{0000-0002-2188-5667},\authornumber{10}
Will Handley\,\orcidlink{0000-0002-5866-0445}\authornumber{11}\separator\authornumber{12}
Edward Higson\,\orcidlink{0000-0001-8383-4614},\authornumber{13}
Michael Hobson\,\orcidlink{0000-0002-0384-0182},\authornumber{12}
Anthony Lasenby\,\orcidlink{0000-0002-8208-6332},\authornumber{11}\separator\authornumber{12}
David Parkinson\,\orcidlink{0000-0002-7464-2351},\authornumber{14}
Livia B. P\'{a}rtay\,\orcidlink{0000-0003-3249-3586},\authornumber{15}
Matthew Pitkin\,\orcidlink{0000-0003-4548-526X},\authornumber{16}
Doris Schneider\,\orcidlink{0000-0001-8284-3324},\authornumber{17}
Leah South\,\orcidlink{0000-0002-5646-2963},\authornumber{18}
Joshua S. Speagle\,\orcidlink{0000-0003-2573-9832} (\begin{CJK*}{UTF8}{gbsn}沈佳士\end{CJK*}),\authornumber{19}\separator\authornumber{20}\separator\authornumber{21}
John Veitch\,\orcidlink{0000-0002-6508-0713},\authornumber{22}
Philipp Wacker\,\orcidlink{0000-0001-8718-4313},\authornumber{17}
David J Wales\,\orcidlink{0000-0002-3555-6645},\authornumber{23}
David Yallup\,\orcidlink{0000-0003-4716-5817}\authornumber{11}\separator\authornumber{12}}
\date{}

\twocolumn[
\maketitle
\begin{abstract}%
We review Skilling's nested sampling (NS) algorithm for Bayesian inference and more broadly multi-dimensional integration. After recapitulating the principles of NS, we survey developments in implementing efficient NS algorithms in practice in high-dimensions, including methods for sampling from the so-called  \gls{constrained prior}. We outline the ways in which NS may be applied and describe the application of NS in three scientific fields in which the algorithm has proved to be useful: cosmology, gravitational-wave astronomy, and materials science. We close by making recommendations for best practice when using NS and by summarizing potential limitations and optimizations of NS.
\end{abstract}
]
\begin{addressbox}%
\addaddress{School of Physics and Astronomy, Monash University, VIC 3800, Australia}\\
\addaddress{Department of Physics, Royal Holloway, University of London, TW20 0EX, United Kingdom}\\
\addaddress{Center for Materials Physics and Technology, U. S. Naval Research Laboratory, Washington, DC 20375, USA}\\
\addaddress{Max Planck Institute for Extraterrestrial Physics, Giessenbachstrasse, 85741 Garching, Germany}\\
\addaddress{Department of Computer Science, University of Bath, Bath, BA2 7PB, United Kingdom}\\
\addaddress{Engineering Laboratory, University of Cambridge, Trumpington Street, Cambridge CB2 1PZ, United Kingdom}\\
\addaddress{Independent researcher}\\
\addaddress{Department of Physics and Institute of Theoretical Physics, Nanjing Normal University, Nanjing, Jiangsu 210023, China. \email{andrew.j.fowlie@NJNU.edu.cn}}\\
\addaddress{Concr Ltd, Babraham Hall House, Cambridge, CB22 3AT}\\
\addaddress{Microscopic Image Analysis Group, Jena University Hospital, Jena, Germany}\\
\addaddress{Astrophysics Group, Cavendish Laboratory, J.J.~Thomson Avenue, Cambridge, CB3~0HE, United Kingdom}\\
\addaddress{Kavli Institute for Cosmology, Cambridge, CB3 0HA, United Kingdom}\\
\addaddress{The D. E. Shaw Group, 1166 Avenue of the Americas, New York, NY 10036, USA}
\addaddress{Korea Astronomy and Space Science Institute, Yuseong-gu, Daedeok-daero 776, Daejeon 34055, Korea}\\
\addaddress{Department of Chemistry, University of Warwick, Coventry, CV4 7AL, United Kingdom}\\
\addaddress{Department of Physics, Lancaster University, Lancaster, LA1 4YB, United Kingdom}\\
\addaddress{Department of Mathematics, Friedrich-Alexander-Universität Erlangen-Nürnberg, 91058 Erlangen, Germany}\\
\addaddress{School of Mathematical Sciences, Queensland University of Technology, Brisbane, QLD 4000, Australia}\\
\addaddress{Dunlap Institute for Astronomy and Astrophysics, University of Toronto, 50 St.\ George Street, Toronto, ON M5S 3H4, Canada}\\
\addaddress{David A. Dunlap Department of Astronomy \& Astrophysics, University of Toronto, 50 St.\ George Street, Toronto ON M5S 3H4, Canada}\\
\addaddress{Department of Statistical Sciences, University of Toronto, 100 St.\ George St, Toronto, ON M5S 3G3, Canada}\\
\addaddress{School of Physics and Astronomy, University of Glasgow, Glasgow, G12 8QQ, United Kingdom}\\
\addaddress{Yusuf Hamied Department of Chemistry, Lensfield Road, Cambridge, CB2 1EW, United Kingdom}
\end{addressbox}

\section{Introduction}\label{sec:intro}
The nested sampling algorithm (NS; \cite{Skilling:2004:nested,Skilling:2006:nested}) was introduced by Skilling in 2004 in the context of Bayesian inference and computation (described in \cref{box:bayes}). The NS algorithm solves otherwise challenging high-dimensional integrals by evolving a collection of live points through the parameter space. The algorithm was immediately adopted in cosmology, owing to the fact that it partially overcomes three major difficulties in the traditional algorithm for Bayesian computation, \gls{Markov chain Monte Carlo}. First, it simultaneously returns results for model comparison and parameter inference. Second, it is successful in \gls{multi-modal} problems. Third, it is naturally self-tuning, permitting it to be applied immediately to new problems. In the 15 years since, the theoretical properties of the algorithm and connections to other computational methods have been partially clarified, and efficient implementations, variants and cross-checks of NS have been developed. The range of applications now extends beyond cosmology and into many other branches of science.

\begin{boxedtextlhs}[box:bayes]{Bayesian inference}
Although NS is a general purpose algorithm for integration, its major application has been integrals in Bayesian inference, and we describe NS using that language. In Bayesian inference\cite{vandeSchoot2021,giulio2003bayesian,gregory2005bayesian,sivia2006data,Trotta:2008qt,von2014bayesian,bailer2017practical} our state of knowledge is quantified by probability and we learn from data by updating probabilities using Bayes' theorem,
\begin{equation*}
\CondProb{A}{B} = \frac{\CondProb{B}{A} \, \Prob{A}}{\Prob{B}}.
\end{equation*}
To use this to learn from data about a model and its parameters, we write it as
\begin{equation*}
\post(\params) = \frac{\like(\params) \, \prior(\params)}{\ev},
\end{equation*}
where the \gls{prior}, $\prior(\params) \equiv \Prob{\params}$ represents what was known about a model's parameters before seeing the data and the \gls{posterior}, $\post(\params) \equiv \CondProb{\params}{D}$, represents what is known after learning from the data. The observed data, $D$, was encoded into the \gls{likelihood} function, $\like(\params) \equiv \CondProb{D}{\params}$. 

The denominator, $\ev \equiv \Prob{D}$, is the \emph{\gls{evidence}} value that appears in \gls{Bayesian model comparison}~\cite{Robert:1995oiy}. It may be written,
\begin{equation*}
\ev = \int \like(\params) \, \prior(\params) \, \diff \params,
\end{equation*}
and so is also known as the marginal \gls{likelihood} and as the normalizing constant, since it normalizes the \gls{posterior} such that $\int \post(\params) \, \diff\params = 1$. The ratio of evidences computed for different models is known as a Bayes factor,
\begin{equation*}
B_{10} = \frac{\ev_1}{\ev_0}.
\end{equation*}
The Bayes factor tells us how we must update the relative plausibility of two models in light of data.
\end{boxedtextlhs}

Here the review of those developments and applications is structured as follows. Below in \nameref{sec:intro}, we recapitulate the origins and principles of NS. We summarize implementations and variants of the NS algorithm, including the developments since its inception, in \nameref{sec:experimentation} and results from NS in \nameref{sec:results}. We describe scientific applications from cosmology, gravitational-wave astronomy, particle physics and materials science in \nameref{sec:applications}. We outline best practices when using NS, including \nameref{sec:repro} and discuss issues with the
technique in \nameref{sec:limitations}. We close by looking forward to the future of NS and Bayesian computation in \nameref{sec:outlook}. Further details are presented in \hyperlink{target:supp}{Supplementary Information}, including a glossary in \cref{sec:gloss} and a simple numerical example in \cref{sec:example}.

\subsection{Multi-dimensional integrals}

Since NS is primarily an algorithm for integration, let us write a general multi-dimensional integral of a function $\like$ over parameters $\params$ as
\begin{equation}\label{eq:Z}
  \ev = \int \like(\params) \, \diff \mu(\params).
\end{equation}
In many scientific problems we need to be able to integrate in high dimensions and for challenging \glspl{integrand}. We assume that the \gls{integrand} is positive, $\like(\params) \ge 0$.

Often, $\ev$ may be a physical quantity such as the total mass of an object distributed with density $\like$ across volumes $\diff \mu(\params)$. Whilst NS is a general method for integration, for concreteness, we view all such applications through the lens of Bayesian inference (see \cref{box:bayes}), with $\diff \mu(\params) \equiv \prior(\params) \diff \params$ seen as an element of \emph{prior probability} with $\prior$ the \emph{\gls{prior}}, normalized by its nature to $\int \prior(\params) \diff \params = 1$. The \gls{integrand}, $\like$, is the modulating \emph{\gls{likelihood}} function (hence the symbol) and $\ev$ is the \emph{\gls{evidence}}. In scientific inference problems, the integral could be over tens if not hundreds of parameters, required to model fundamental effects as well as the calibration and systematics of complicated experimental measurements~\cite{2020arXiv201209874A}.

We may rewrite the \gls{integrand} through the elementary factorization known as Bayes' theorem,
\begin{equation}
\like(\params) \times \pi(\params) = \ev \times \post(\params),
\label{eq:Bayes}
\end{equation}
where
\begin{equation}\label{eq:P}
  \post(\params) = \frac{\like(\params) \prior(\params)}{\ev},
\end{equation}
is the \emph{\gls{posterior}}, normalized to $\int \post(\params) \, \diff\params = 1$. Notation apart, however, all we are doing here is decomposing the \gls{integrand} into a magnitude $\ev$ and a shape $\post(\params)$.

Historically, Bayesian computation focused on only the shape $\post(\params)$, partly owing to controversies around \gls{Bayesian model comparison} such as its sensitivity to the choices of \gls{prior}~\cite{Robert:1995oiy}, and partly due to computational difficulties~\cite{2020arXiv200406425M}. However, shapes and magnitudes both matter, especially in the general setting of multi-dimensional integration beyond \gls{Bayesian model comparison}.
NS \cite{Skilling:2004:nested, Skilling:2006:nested} surmounts the challenge by computing shapes and magnitudes simultaneously.

\subsection{Simplifying multi-dimensional integrals}

Before introducing NS, let us attempt to simplify the general integral in \cref{eq:Z}. Consider traditional Riemann-style integration. This decomposes the space into volume elements $\Delta\params$, typically small cubes, and performs a sum over them. Small cubes, however, rapidly become infeasible in multi-dimensional integration because their cost grows exponentially with dimension --- this is the ``\gls{cod}.''

\begin{figure*}[thb!]
\begin{subfigure}[b]{0.5\textwidth}
\centering
\includegraphics[width=0.9\linewidth]{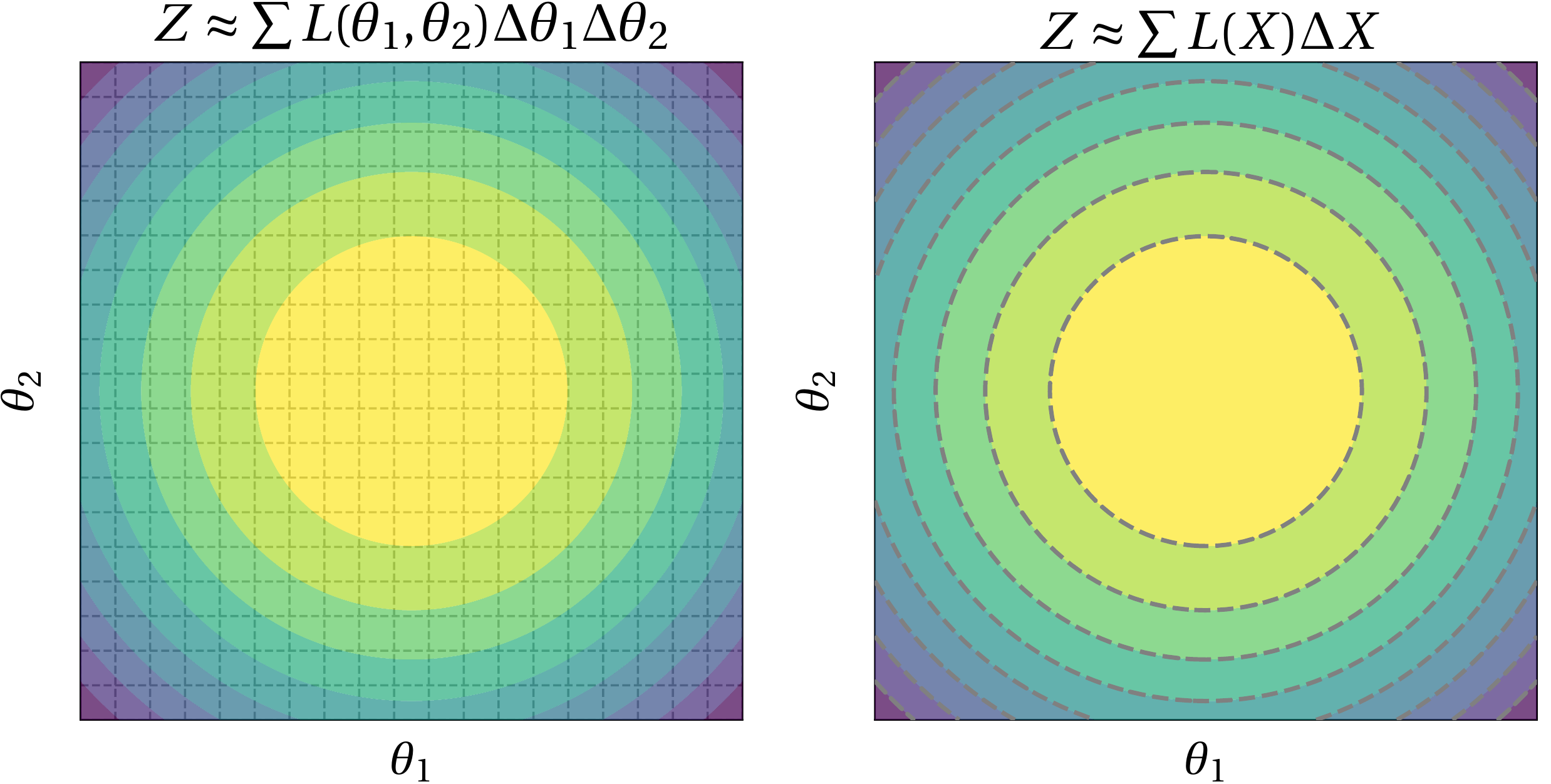}
\caption{\captiontitle{The NS \gls{evidence} identity} The colours represent contours of a two-dimensional \gls{likelihood} function. Rather than summing over little cubes (left), we combine cubes of similar \gls{likelihood} together and sum over them (right).}
\label{fig:combine_cells}
\includegraphics[width=0.9\linewidth]{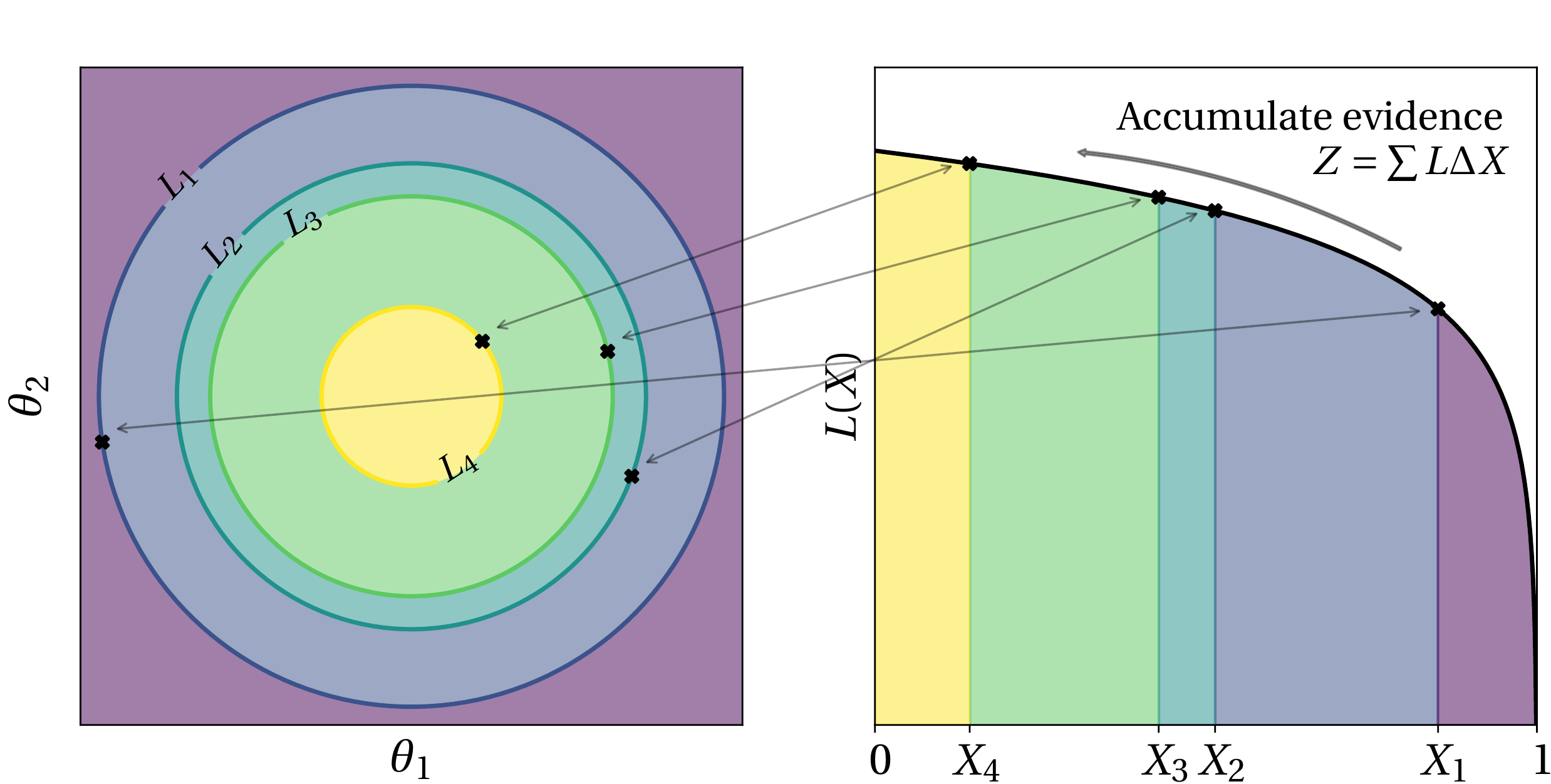}
\caption{\captiontitle{NS on a two dimensional problem} We show the dead points and their \glspl{iso-likelihood contour} (left) and the corresponding contributions to the \gls{evidence} integral (right). The volumes $X_i$ are estimated statistically in NS.}
\label{fig:ns_2d}
\end{subfigure}
\begin{subfigure}[b]{0.5\textwidth}
\centering
\includegraphics[width=0.9\linewidth]{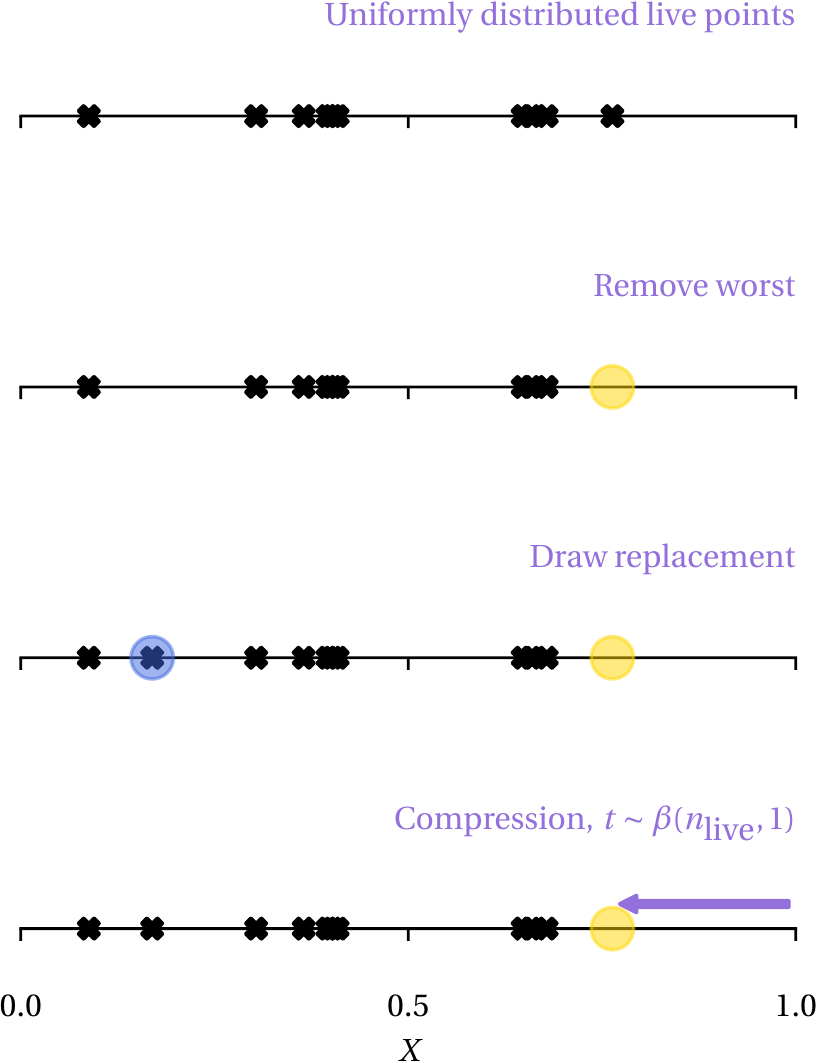}\vspace{0.5cm}
\caption{\captiontitle{Compression in one iterate of NS}}
\label{fig:compression}
\end{subfigure}
\caption{\captiontitle{Illustrations of NS algorithm}}
\label{fig:ns}
\end{figure*}

We don't, however, need to decompose our space into little cubes; our cells can be any shape we want. The integrals needed for quantification,
\begin{equation}
  \ev = \int \like(\params) \prior(\params) \, \diff\params = \lim\limits_{|\Delta\params| \to 0} \, \sum \like(\params) \prior(\params) \, \Delta\params,
\end{equation}
are defined as limiting sums over volume elements which should be small enough to keep $\post(\params)$ almost constant \emph{regardless of shape}. We may therefore combine the cells in which the \gls{integrand} is almost constant. Schematically, we may write
\begin{equation}\label{eq:schematic_Z}
  \ev = \sum \like(X) \Delta X,
\end{equation}
where $\Delta X$ is the volume of cells that share \gls{likelihood} $\like(X)$ weighted by the \gls{prior} $\prior(\params)$. This is illustrated schematically in \cref{fig:combine_cells} and works whether the \gls{integrand} is \glssymbol{uni-modal} or \gls{multi-modal}.

\begin{boxedtextlhs}[box:math]{Mathematical details}
The idea of NS is to transform \cref{eq:Z} into \cref{eq:riemann} which can be approximated more efficiently using the described Monte Carlo approach with active samples:
\begin{equation*}
  \ev = \int_\Omega  \like(\params)  \diff\mu(\params) =\int_0^1 \tilde{\like}(X) \diff X,
\end{equation*}
where $\Omega$ is the \gls{parameter domain} and $\tilde{\like}$ is an \gls{overloaded} form of $\like$, as described next. To achieve this transformation, we define the \gls{survival function} $X: \mathbb{R} \to [0,1]$, $X(\lambda) = \mu(\{z\in\Omega:  \like(z) > \lambda\})$, that is the $\mu$-\gls{measure} of the $\lambda$-\gls{super-level-sets}. Then, we introduce a mapping $\Phi : \Omega \to [0,1]$ with
\begin{equation*}
  \Phi(\params) = X({\like}(\params)) = \mu\left( \left\lbrace z \in \Omega : {\like}(z) > {\like}(\params) \right\rbrace  \right)
\end{equation*}
This mapping $\Phi$ is the transformation which allows us to shift the integration from $\Omega$ to $[0,1]$, by virtue of the \gls{push-forward measure} $\mu \circ \Phi^{-1}$ on $[0,1]$. Now, we can define
\begin{equation*}
  \tilde \like: [0,1] \to \mathbb{R}, \quad \tilde{\like}(\xi) = \sup \left\{\lambda \in \operatorname{Im}  \like : X(\lambda) > \xi \right\}.
\end{equation*}
Thus the integral transformation above which provides an alternative characterization of $\ev$ is true because
\begin{align*}
\int_\Omega \like(\params) \, \diff \mu(\params)
&{=} \int_\Omega\tilde{\like} \left( X\left({\like}(\params)\right)\right) \, \diff \mu(\params) \\
&{=} \int_\Omega (\tilde{\like} \circ \Phi)(\params)  \, \diff \mu(\params)
  =  \int_{[0,1]} \tilde{\like}(x) \, \diff (\mu\circ \Phi^{-1})(x)\\
&{=} \int_0^1 \tilde{\like}(x) \, \diff x
\end{align*}
Here, $\tilde{\like}$ is a generalized inverse of $X$ under suitable assumptions on $\like$, see \ccite{embrechts2013note,de2015study}. 
This integral transformation holds at least in the case that $\like$ has no plateaus of positive \gls{prior} \gls{measure}, owing to the fact that in this case $\mu\circ \Phi^{-1}$ is indeed the uniform \gls{measure} on $[0,1]$, that is $\diff \mu\circ \Phi^{-1}(x) = \diff x$.

If $\like$ has a plateau of non-negligible mass, i.e., there exists a level $\lambda^\star$ such that $\mu(\{z \in \Omega: \like(z) = \lambda^\star\}) > 0$, then the derivation is more challenging, see also \nameref{sec:limitations}.
\end{boxedtextlhs}

We can reach \cref{eq:schematic_Z} more concretely by noting that the \gls{evidence} is the expectation of a non-negative random variable, such that it may be written as
\begin{equation}\label{eq:lebesgue}
  \ev = \int X(\like) \, \diff\like,
\end{equation}
where the volume variable $X$,
\begin{equation}\label{eq:X}
 X(\threshold) = \int\limits_{\like > \threshold} \prior(\params)  \diff\params, %
\end{equation}
is the volume enclosed by contour $\threshold$. This result can be readily proven by integration by parts (see also sec.~21 in \ccite{billingsley}). Applying integration by parts again to \cref{eq:lebesgue}, we obtain the familiar NS \gls{evidence} identity,
\begin{equation}\label{eq:riemann}
  \ev = \int_0^1 \like(X) \, \diff X,
\end{equation}
providing that $\like(X)$, the inverse of $X(\lambda)$, indeed exists and that the \gls{evidence} is finite. This formalizes the schematic \cref{eq:schematic_Z}. We discuss this result more formally in \cref{box:math}.


\subsection{Nested sampling}\label{sec:NS}

As 
the multi-dimensional integral in \cref{eq:X} is impractical in high dimension, some sort of statistical estimation is inevitable.
NS starts with an ensemble of $\nlive$ random locations $\params$ drawn from the \gls{prior}, $\prior(\params)$, each of which has its \gls{likelihood} $\like(\params)$ which we can place in ascending order.
Crudely, if we discarded the lowest half of the values, the survivors would be random samples taken within the restricted volume $\like > \text{Median}[\like]$, which would statistically be roughly half the original volume. This allows us to make a statistical estimate of the volume variable in \cref{eq:X}. Repeating that $\niter$ times would yield compression by a factor of about $2^{\niter}$. This is the exponential behaviour required to overcome the \gls{cod}.


Thus NS works by statistical estimates of the compression, which is a general and fundamental operation that can be used in various ways not limited to those in \cref{tab:applications}. The \gls{evidence} identity in \cref{eq:lebesgue} isn't required in every application; many applications only use the compression in \cref{eq:X}. We present scientific applications in \nameref{sec:applications}.

\begin{table*}[thb!]
\centering
\renewcommand{\arraystretch}{1.25}
\begin{tabularx}{\textwidth}{l X}
\toprule
NS application & Details\\
\midrule
Integration & Perform the general multi-dimensional integral \cref{eq:Z} for positive integrands.\\
Global optimization & Maximize the \gls{likelihood}, $\like$, by compressing to $\hat\params$, the maximum of $\like(\params)$, with no restriction to \gls{uni-modal} distributions. This may require strict settings and be more computationally expensive than integration; see \ccitetable{Akrami:2009hp,Feroz:2011bj} for further discussion.\\
Bayesian inference & NS simultaneously computes the \gls{posterior} and the Bayesian \gls{evidence}, allowing parameter inference and model comparison.\\
Approximate Bayesian Computation & Perform efficient Approximate Bayesian Computation by applying NS to the joint space of parameters and data~\citetable{Brewer:2016scw}\\
Statistical thermodynamics & If we use the Boltzmann factor as the \gls{likelihood}, i.e., $\like(\params) = e^{-\beta E(\params)}$, where $E$ is the energy of the state $\params$ among $N$, we may use NS to compute the \gls{partition function}.
By accumulating for several temperatures $T$ in parallel, one can plot thermodynamic functions such as the specific heat $C_V = T\,\diff S/ \diff T$ as functions of temperature without needing multiple runs.\\
Rare event sampling & The volume variable $X$ may be interpreted as the probability of a rare event~\citetable{walter2015point,birge2013split} or used to compute a $p$-value in frequentist statistics~\citetable{Fowlie:2021gmr}.\\
\bottomrule
\end{tabularx}
\caption{\captiontitle{Applications of NS}}
\label{tab:applications}
\end{table*}

\subsection{Formulation} \label{sec:Formulation}

We now present the NS algorithm in more detail. We assume that there are no regions of constant \gls{likelihood} resulting in \gls{likelihood} plateaus (see \nameref{sec:limitations} for further discussion). The NS algorithm begins by drawing an ensemble of $\nlive$ samples from the \gls{prior}. We compute the \gls{likelihood} for each sample. We denote the smallest \gls{likelihood} by $\threshold$ and we discard that point. The remaining live points are now distributed over a compressed volume; we denote the factor by which the volume compressed by $\compression$. Finally, a replacement point is drawn from the \gls{prior} subject to $\like > \threshold$, that is, from the  \gls{constrained prior},
\begin{equation}\label{eq:constrained_prior}
\prior^\star (\params) \propto
    \begin{cases}
        \prior(\params) & \text{if } \like(\params) > \threshold \\
        0 &\text{otherwise}.
    \end{cases}
\end{equation}
This leaves a new ensemble with $\nlive$ samples obeying a \gls{likelihood} constraint $\like > \threshold$.

As they are drawn from the \gls{constrained prior}, the volumes $X$ associated with the live points are uniformly distributed. Thus the compression associated with the discarded outermost sample, $\compression$, corresponds to the smallest of $\nlive$ uniform random variables. This follows a $\betadist(\nlive, 1)$ distribution,
\begin{equation}
\post(\compression) = \nlive \compression^{\nlive - 1}.
\label{eq:beta}
\end{equation}
The first factor accounts for the fact that any live point could be the outermost and the second factor for the fact that $\nlive - 1$ uniformly distributed samples lie above the outermost sample at $\compression$ (the remaining sample lies at $\compression$).

We started from unconstrained samples drawn from the full original volume $X_0=1$. As we repeat this process of replacing the outermost points, the successive compressions by factors $\compression_1, \compression_2, \compression_3,\dots$ lead to exponentially decreasing inferred volumes $X_1, X_2, X_3, \dots$
\begin{equation}
    X_0 = 1. \quad X_{i + 1} = \compression_{i + 1} X_i.
\end{equation}
We illustrate a single iteration of NS in \cref{fig:compression}.
If we are most interested in the magnitude of the \gls{evidence}, $\log \ev$, we should consider,
\begin{equation}\label{eq:geometric}
\log t \approx \langle \log \compression \rangle = -\frac{1}{\nlive},
\end{equation}
as under repeated multiplication $X_k = \compression_1 \compression_2 \dots \compression_{k}$ it's the \emph{logarithms} that add. For a more complete inference, the compression factors $\compression$ may be sampled directly from $\betadist(\nlive, 1)$. See \cref{sec:compression} for further discussion. NS therefore estimates volumes through probability not geometry, topology, or even dimension.
We do not get a definite compression value, only a distribution of what it might have been.

Having obtained estimates of the volume $X(\threshold)$ at each of $\niter$ iterations, we may accumulate the  \gls{evidence} via \cref{eq:lebesgue} or \cref{eq:riemann}, for example by the trapezium rule,
\begin{equation}\label{eq:sum_Z}
    \ev = \sum_{i = 1}^{\niter} w_i \threshold_i,
\end{equation}
where the weights are
\begin{equation}\label{eq:weights_Z}
    w_i = \frac12 \left(X_{i-1} - X_{i+1}\right).
\end{equation}
The sum in \cref{eq:sum_Z} converges to the desired integral~\cite{Chopin_2010,2009AIPC.1193..277S,evans2007discussion}. This is the magnitude; we obtain shape by assigning weights to each sample (see \ccite{Chopin_2010} for discussion),
\begin{equation}\label{eq:posterior_weights}
    \post_i = \frac{w_i \threshold_i}{\ev},
\end{equation}
normalized such that $\sum \post_i = 1$. These are the \gls{posterior} weights of the dead points; the shape may be recovered by for example a weighted histogram or other density estimation methods.  We show the whole algorithm schematically in \cref{alg:ns} and the summation for a two-dimensional problem in \cref{fig:ns_2d}.

\begin{algorithm*}[thb!]
\caption{\captiontitle{Schematic of the NS algorithm for the general multi-dimensional integral in \cref{eq:Z}} Techniques for drawing replacements and stopping criteria are discussed in \nameref{sec:exploration} and \nameref{sec:stopping} respectively.}\label{alg:ns}
\SetAlgoLined
 Choose an estimate of the compression factor, e.g., $\compression = e^{-1/ \nlive}$\;
  Initialize volume, $X = 1$\, and integral, $\ev = 0$\;
  Sample $\nlive$ points from the \gls{prior} --- the live points\;
 \Repeat{stopping criteria satisfied}{%
    Let $\threshold$ be the minimum $\like$ of the live points\;
    Replace live point corresponding to $\threshold$ by one drawn from the \gls{prior} subject to $\like > \threshold$\;
    Increment the estimate of the integral, $\ev = \ev + \threshold \Delta X$, with e.g., $ \Delta X = (1 - \compression) X$\;
    Contract volume, $X = \compression X$\;
 }
 Add estimate of remaining  \gls{evidence}, e.g., $\ev = \ev + \bar\like X$\, where $\bar\like$ is the average \gls{likelihood} among the live points\;
 \KwRet{Estimate of integral, $\ev$}
\end{algorithm*}

Compared to the sketch of NS in \cref{sec:NS}, we replace a single live point per iteration, rather than half of the live points. Whilst the number of replacements can be varied, one replacement is optimal (though see considerations in \nameref{sec:parallel}). This is because for $r \ll \nlive$, replacing $r$ points per iteration would reach the \gls{posterior} \gls{bulk} in about $r$ times fewer iterations. While the computational expense wouldn't change as $r$ replacements are required per iteration, the error estimates would scale as $\sqrt{r}$ because
reducing the number of iterations increases the relative Poisson noise in the number of iterations.

In the last decade or so, analogies between NS and statistical mechanics (see \cref{box:stat_mech}) and other statistical methods, including sequential Monte Carlo~(SMC; \cite{Salomone2018}) and subset simulation in rare event sampling~\cite{Au_2001,beck2015rare,walter2015point,birge2013split}, among others~\cite{10.1093/imamat/26.2.151,doi:10.1063/1.1701695,NEURIPS2021_8dd291cb,PhysRevLett.122.150602,polson2015verticallikelihood}, have been recognized. The connections to SMC and 
an SMC variant of NS are discussed in \cref{box:smc}. 
There are, of course, other strategies for computing the evidence; see \ccite{Robert2009,Knuth2015,Zhao2016,llorente2020marginal} for reviews. Notable examples include approximating the integrand by a tractable function\cite{tierney1986}, Chib's method using density estimation and Gibbs' sampling\cite{chib}, importance sampling\cite{10.2307/1913641}, and techniques that re-use \gls{Markov chain Monte Carlo} draws\cite{newton1994approximate}.
Broadly speaking, NS lies in a class of algorithms that form a path of bridging distributions, and evolves samples along that path~\cite{10.1214/ss/1028905934,10.1214/13-STS465}. NS stands out because the path is automatic and smooth --- we compress in $\log X$ by on average $1 / \nlive$ at each iteration --- and because along the path we compress through constrained \glspl{prior}, rather than from the \gls{prior} to the \gls{posterior}. This was in fact a motivation for NS as it avoids phase transitions --- abrupt changes in the bridging distributions --- that cause problems for other methods including path samplers such as annealing. We further discuss NS's historical background and contrast it with annealing in \cref{app:annealing}. 

\subsection{Uncertainties}\label{sec:uncertainty}

Our estimates of the magnitude and shape in \cref{eq:Z} must be accompanied by a discussion of the bias and statistical uncertainty. The latter originates from our noisy estimates of the compression factors. We may estimate the resulting statistical uncertainty in the  \gls{evidence} by considering the compression required to reach the \gls{bulk} of the \gls{posterior}. This may be quantified by the information content~\cite{6773024,4082152}
\begin{equation}\label{eq:kl}
  \kl = \int \post(\params) \log \left(\frac{\post(\params)}{\prior(\params)}\right) \, \diff\params
\end{equation}
known in statistics as the \gls{Kullback-Leibler divergence}. We can write it using the volume variable as
\begin{align}
  \kl &= \int \post(X) \log \post(X) \, \diff X \\
      &= - \int \post(X) \log X \, \diff X + \int \post(\log X) \log \post(\log X) \, \diff \log X\label{eq:kl_as_function_of_volume}
\end{align}
where $\post(X) \equiv \like(X) / \ev$ is the posterior density of the volume. In \cref{eq:kl_as_function_of_volume} we see that the KL divergence equals minus the posterior expectation of $\log X$ minus the differential entropy associated with the posterior of $\log X$. As the first term typically dominates, the KL divergence provides a measure of compression. 

Thus from \cref{eq:geometric} it's likely to take about
\begin{equation}\label{eq:runtime}
\nbulk \simeq \nlive \kl
\end{equation}
iterations to compress to the \gls{bulk} of the \gls{posterior} at $\log X = -\kl$, and this count is likely to be subject to $\sqrt{\nbulk}$ Poisson variability. Neglecting contributions from outside the bulk, we may write the evidence sum in \cref{eq:sum_Z} as
\begin{equation}
    \ev \simeq e^{-\nbulk / \nlive} \sum_{i = \nbulk}^{\niter} (1 - t)^{i - \nbulk} \threshold_i.
\end{equation}
We see that the final estimate of $\log \ev$ will be plausibly subject to an approximately Gaussian uncertainty from the first factor,
\begin{equation}\label{eq:delta_Z}
\Delta\log\ev \approx \sqrt{\frac{\kl}{\nlive}}.
\end{equation}
Thus NS statistical uncertainties scale as $1 / \sqrt{\nlive}$ as usual for statistical uncertainties (see \ccite{Chopin_2010} for an alternative proof and discussion).

Although NS was first introduced with this estimate~\cite{Skilling:2004:nested}, it can be unreliable. Authority rests with repeated simulation through \cref{eq:beta} of what the compressions might actually have been. See \ccite{2011MNRAS.414.1418K} for further discussion of the statistical uncertainty in the NS estimates and \cref{sec:uncertainty_posterior} for discussion of uncertainties in NS estimates of the posterior. As well as this statistical uncertainty, there are four potential sources of bias: bias originating from failure to faithfully sample from the  \gls{constrained prior}, bias originating from the choice of estimator for the compression factor, the generally negligible quadrature error, and the potentially important truncation error in \cref{eq:sum_Z}. The latter occurs as we stop after a finite number of iterations and is further discussed in \nameref{sec:stopping}. Provided that NS is appropriately configured, the statistical uncertainty usually dominates.

We see that difficulty in NS does not in fact lie in dimension but in compression from the \gls{prior} to the \gls{posterior}: the compression and the resolution $\nlive$ alone determine the uncertainty and the run-time. To maintain a given uncertainty, by \cref{eq:delta_Z} we require $\nlive \propto \kl$ live points and by \cref{eq:runtime} $\niter \propto \kl^2$ iterations. If the \gls{prior} and \gls{posterior} are factorizable into a term for each dimension, by \cref{eq:kl} the \gls{Kullback-Leibler divergence} is additive, and so scales linearly with dimension, $\dimension$, and so run-time goes like $\order{\dimension^2}$. NS beats the exponential scaling with dimension expected from the \gls{cod}.

\begin{boxedtextrhs}[box:stat_mech]{Statistical mechanics analogy}
There is a strong analogy between Bayesian inference and statistical mechanics. This suggests that NS might be useful in exploring problems that are typically the subject of statistical mechanical analysis.
Consider $\params$ a \gls{microstate} and \(E(\params) = -\log \like(\params)\) its energy.
Then the \emph{\gls{microcanonical ensemble}} includes all states with \(E(\params) = \epsilon\) where \(\epsilon\) is a constant energy level.
The volume of state space corresponding to a given energy $\epsilon$ is given by the density of states,
\begin{equation*}
g(\epsilon) = \int \delta(E(\params) - \epsilon)\prior(\params) \, \diff \params
\end{equation*}
and taking the \gls{prior} to be uniform corresponds to the {\em ergodic hypothesis}, namely that each \gls{microstate} that is allowed by applicable conservation laws is equally likely to be observed. The Laplace transform of the density of states is the canonical \gls{partition function} $\ev(\beta) = \int e^{-\beta E} g(E) \diff E$, which corresponds to the generalized  \gls{evidence} (\cref{eq:generalised-evidence}). The canonical ensemble is an alternative description of thermodynamic states that is based on the {\em inverse temperature} $\beta$ rather than the energy level $\epsilon$. In the canonical ensemble, states follow the Boltzmann distribution $p(\params\mid\beta) = \exp\{-\beta E(\params)\} / \ev(\beta)$. NS in essence tracks the cumulative density of states via:
\begin{equation*}
  X(\like) = \int\limits_{E \le -\log \like} g(E) \, \diff E,
\end{equation*}
The practical approximation corresponding to \cref{eq:sum_Z,eq:weights_Z} is
\begin{equation*}
\ev(\beta) = \sum {e^{-\beta E_i} \left(X_{i-1}  -  X_{i+1} \right)/2}.
\end{equation*}

During NS, states are generated from the \gls{prior} constrained by an upper energy limit \(\epsilon = -\log \threshold\), which can be achieved with any number of techniques, such as simple rejection sampling, Galilean Monte Carlo and Hamiltonian dynamics\cite{ConPresNS} or Creutz' microcanonical demon algorithm~\cite{creutz1983microcanonical, habeck2015nested}.
The information entropy of the  \gls{constrained prior} is the {\em volume entropy} \(\log X(\epsilon)\) (also known as Gibbs entropy).
As NS progresses from one energy limit \(\epsilon\) to the next \(\epsilon' < \epsilon\), the volume entropy changes by \(\Delta H = \log[X(\epsilon) / X(\epsilon')]\) at a rate that is constant on average: $\langle \Delta H \rangle = 1 / \nlive$.

Widely-used tempering methods such as simulated annealing work in the canonical ensemble and use the inverse temperature $\beta$ as a control parameter to weight each \glspl{microstate}.
Alternatives include the Wang-Landau method\cite{WangLandau} and NS, both of which use energy as a control parameter.
In contrast to the ensemble property \(\beta\), a key advantage of \(E(\params)\) is that it can be evaluated for a single \gls{microstate}.
The Wang-Landau method uses a fixed, predefined set of energy bins, while NS
constructs
a sequence of energy levels at runtime.
The sequence is optimal in that it achieves constant thermodynamic speed because changes in volume entropy are constant on average.
Therefore, NS elegantly
avoids a major problem of canonical annealing methods: designing a good temperature schedule.
\end{boxedtextrhs}

\section{Experimentation}\label{sec:experimentation}
In this section we will discuss how to implement NS, including considerations such as choosing the number of live points, drawing new live points from the  \gls{constrained prior}, parallelization and deciding when to stop.

\subsection{Choice of the number of live points}\label{sec:choice_n_live}

\begin{figure*}
\begin{subfigure}[c]{\textwidth}
    \centering
    \includegraphics[width=0.9\linewidth]{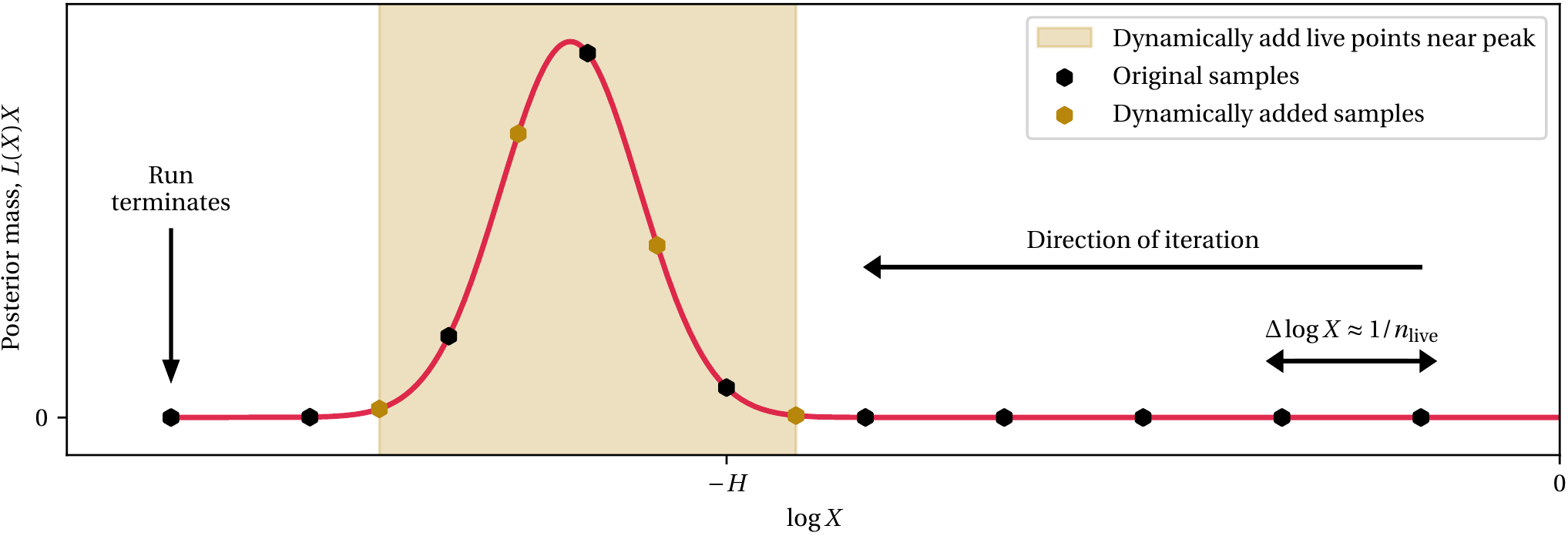}
    \subcaption{\captiontitle{Schematic representation of an NS run} The curve $\like(X) X$ shows the
      relative \gls{posterior} mass, the \gls{bulk} of which lies in a
      tiny fraction $e^{-\kl}$ of the volume. Most of the original samples lie in regions with negligible \gls{posterior} mass. In dynamic NS, we add samples near the peak.}
\label{fig:ns_evidence}
\end{subfigure}
\begin{subfigure}[c]{\textwidth}
    \centering
    \includegraphics[width=0.9\textwidth]{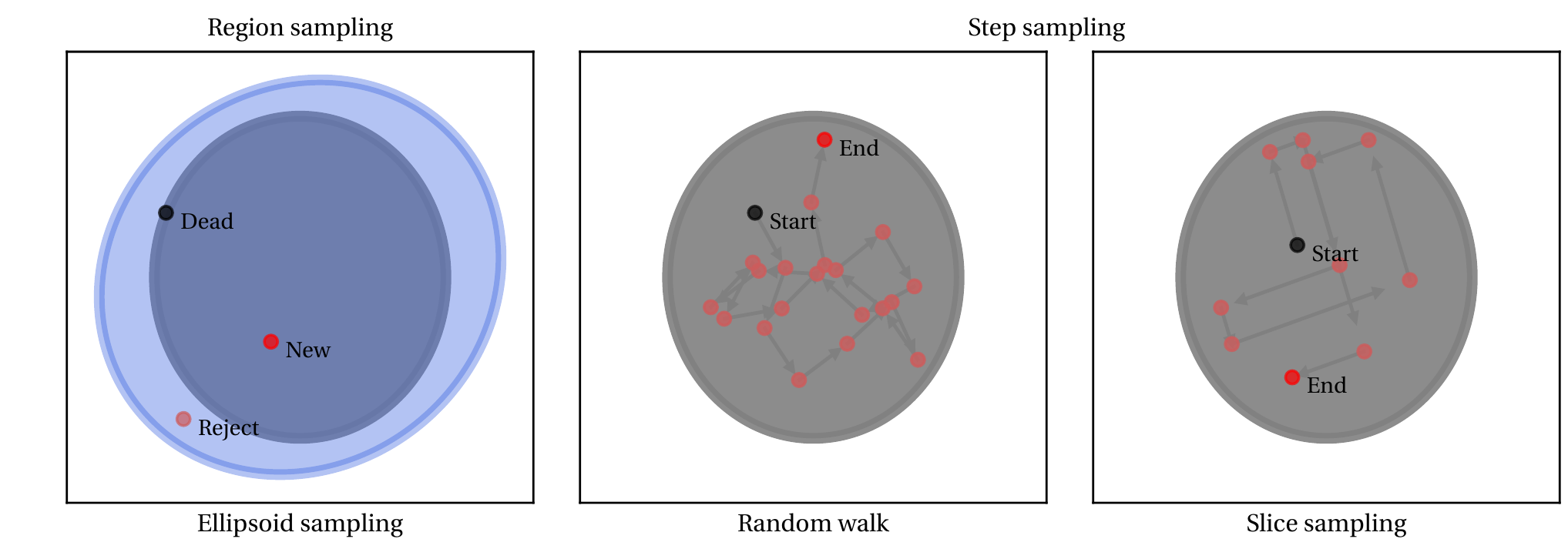}
    \subcaption{\captiontitle{Examples of strategies for sampling from the  \gls{constrained prior}} We must sample from the true \gls{iso-likelihood contour} (grey ellipse). In region sampling (left) we bound the existing live points (blue ellipse) and draw a new sample from within that bound; some proposals may be rejected. In step sampling (right) we select a live point and perform a sequence of steps inside the contour to obtain an independent draw.
    }
    \label{fig:lrps_strategies}
\end{subfigure}
\caption{\captiontitle{Experimentation in NS} \subref{fig:ns_evidence} Illustration of an NS run and \subref{fig:lrps_strategies} strategies for sampling from the  \gls{constrained prior}.}
\label{fig:exp}
\end{figure*}

As discussed in \nameref{sec:Formulation}, the number of live points controls the \emph{rate} of exponential compression during an NS run; we compress inwards by about $\Delta \log X \approx 1 / \nlive$ per iteration. This means that run-time scales as $\order{\nlive}$ (\cref{eq:runtime}) and that the dominant uncertainties on the \gls{evidence} integral (\cref{eq:delta_Z}) and \gls{posterior} scale as $\order{1/\sqrt{\nlive}}$. The above considerations are represented graphically in \cref{fig:ns_evidence}.

There is thus a trade-off between run-time and uncertainty. Despite that, there isn't a
straight-forward method for choosing the number of live points if the required compression isn't known ahead of time.
The number should be chosen, furthermore, bearing in mind the alternative role
that it plays as a resolution parameter for NS\cite{MultiNest1,partay2010efficient}, especially in \gls{multi-modal} problems. In particular, $\nlive$ should be large enough that at any time the \gls{constrained prior} splits into disjoint \glspl{mode}, 
at least one live point lies inside the footprint
of each \gls{mode}. As a rough rule of thumb, if the \gls{constrained prior} occupies a total volume $X$, only \glspl{mode} with a footprint greater than about $X  / \nlive$ may be reliably found\cite{sivia2006data,partay2010efficient}. This defines a resolution down
to which the \gls{posterior} is reliably sampled. \Glspl{mode} with smaller
footprints are typically not located and correctly sampled, and hence
will also not contribute to the \gls{evidence} estimate. Moreover, to sample reliably and efficiently from the \gls{constrained prior}, it is usually advisable that $\nlive$ exceeds the dimensionality of the parameter space. 

\begin{boxedtextlhs}[box:smc]{Sequential Monte Carlo}
Connections between NS and a form of rare-event SMC sampler \cite{Cerou2012} can be exploited to develop an NS-SMC algorithm with desirable theoretical properties \cite{Salomone2018}.

SMC samplers evolve $n$ samples (referred to as ``particles'') through a series of distributions $\prior_t$ for $t=0,\ldots,T$ using reweighting, resampling and mutation steps. Of particular interest is a form of rare-event SMC \cite{Cerou2012} that involves sampling from the sequentially \gls{constrained prior} distribution. This SMC sampler has the sequence of distributions
\begin{align*}
    \eta_t \propto \mathbb{1}(\like(\params)\geq \threshold_t) \prior(\params),
\end{align*}
where $\mathbb{1}(\cdot)$ denotes an \gls{indicator function} and $0=\threshold_0<\cdots<\threshold_t<\cdots <\threshold_{T+1}= \infty$. Each iteration $t$ of the sampler involves sampling $n$ particles with replacement from the current set of points satisfying $\like(\params)\geq \threshold_t$ and then diversifying those particles through several iterations of an $\eta_t$-invariant \gls{Markov chain Monte Carlo} kernel. 

Noticing the similarities between the two algorithms, \ccite{Salomone2018} built upon these rare-event SMC samplers to create the NS-SMC algorithm for evidence and posterior estimation. Approximation of $\ev$ and $\post$ in NS-SMC is done by weighting the particles from iteration $t$ to target a shell of the \gls{posterior},
\begin{align*}
    P_t \propto \mathbb{1}(\threshold_t<\like(\params)\leq\threshold_{t+1}) \like(\params)\prior(\params).
\end{align*}
While original NS and NS-SMC both sample from the sequentially \gls{constrained prior} distribution, there are several key differences. Crucially, NS-SMC uses weights based on importance sampling rather than numerical quadrature, it naturally handles the region beyond the largest threshold with no truncation error and it uses a different sampling mechanism. As a result, NS-SMC avoids the issues with bias in NS. Under mild conditions, including the situation where a finite number of \gls{Markov chain Monte Carlo} steps are used, NS-SMC produces unbiased and consistent estimates of $\ev$ and consistent estimates of $\post$ as $n\rightarrow \infty$ \cite{Salomone2018}. Practically, NS can result in noticeably biased estimates of the \gls{evidence} compared to NS-SMC for the same computational cost when the MCMC kernel is inefficient.
\end{boxedtextlhs}

\subsection{Dynamic nested sampling}\label{sec:dns}

We have so far only considered NS with a fixed number of live points, and noted that the uncertainties in both \gls{posterior} distributions and  \gls{evidence} estimates are reduced by increasing this number.
However, the  \gls{evidence} depends on an accurate estimate of the total compression when we reach the \gls{posterior} \gls{bulk}. The \gls{posterior}, on the other hand, depends only on an accurate estimate of the relative compression once inside the \gls{posterior} \gls{bulk}. The former uncertainty cancels in the \gls{posterior} weights in \cref{eq:posterior_weights}, as they are invariant under rescaling the estimates of the volume variable. This reflects the fact that whereas the  \gls{evidence} may depend strongly on the size of the prior, the \gls{posterior} usually depends only weakly on its shape.

We are thus motivated to consider a \emph{dynamic} number of live points to efficiently reduce uncertainties in parameter inference. As shown in \cref{fig:ns_evidence}, we may quickly compress to the \gls{posterior} \gls{bulk} using few live points. Upon reaching it, we may increase the number of live points, reducing uncertainty in the \gls{bulk} of the \gls{posterior} mass. We denote schemes that vary the number of live points as \emph{dynamic} NS~\cite{higson2019dynamic}. Open-source dynamic NS software packages include
\dynesty{} \cite{2020MNRAS.493.3132S} and \dyPolyChord{} \cite{Higson2018}; see \cref{tab:codes}.
The gains are greatest in problems with substantial compression to the \gls{posterior} \gls{bulk}.

In dynamic NS schemes, the number of live points can be automatically
adjusted to maximize a user-specified objective for a fixed computational budget.
Usually, the run starts with an exploratory NS run using a constant
number of live points. We spend the remaining computational budget by repeatedly increasing the number of live points in
the most important regions of volume judged according to the objective (for example the shaded region in \cref{fig:ns_evidence}). Because dynamic NS re-winds a run and adds extra samples anywhere, running for longer reduces uncertainties and increases the effective sample size, unlike in ordinary NS.

In the original dynamic NS algorithm~\cite{higson2019dynamic}, a user specifies their objective by assigning a relative importance to reducing uncertainties in the \gls{posterior} and the \gls{evidence}.
When focusing on \gls{posterior} inferences, dynamic NS can achieve orders of magnitude reductions in computational cost for a fixed uncertainty.
The approach can also improve \gls{evidence} calculation accuracy for a fixed number of samples, and can improve \gls{posterior} inferences and \gls{evidence} calculations simultaneously.
This objective was generalised by the reactive NS \cite{NSmethodsReview} variant of dynamic NS. This considers the computation as a graph, with the nodes being the live and dead points, and edges indicating the replacement of a point by another. Multiple agents can then add live points (edges) where needed, and optimize towards additional goals, such as the effective sample size, or the number of samples per cluster.
Lastly, we note a significant variant of NS, diffusive NS \cite{brewer2011diffusive}, in which the number of live points at a given \gls{likelihood} threshold can change. In this variant, random walks starting from the existing live points are permitted to step down as well as up in \gls{likelihood}, refining the typical \gls{likelihood} in a volume range $X$.

\subsection{Exploration strategies}\label{sec:exploration}

NS progresses by replacing live points by independent samples drawn from the  \gls{constrained prior} in \cref{eq:constrained_prior}, that is the prior restricted to regions in which the \gls{likelihood} exceeds a threshold. This is the major difficulty in efficiently and reliably implementing NS, especially in \gls{multi-modal} problems. Provided one successfully samples from the constrained prior, however, NS works identically in \gls{uni-modal} and \gls{multi-modal} settings. 

Whilst we could simply sample from the entire prior until we find a sample for which the \gls{likelihood} exceeds the threshold, this rapidly becomes incredibly inefficient  due to the exponential reduction in the volume contained within the  \gls{constrained prior} at every iteration. Fortunately, the current set of live points and the estimate of the volume enclosed by the contour may guide our search for new live points. There are two main classes of methods for sampling from the  \gls{constrained prior}~\cite{Buchner2014test}: region samplers and step samplers. They are illustrated in \cref{fig:lrps_strategies}. Analogous to the choices of \glspl{transition kernel} in \gls{Markov chain Monte Carlo}, the choices here lead to various flavors of NS with different performance characteristics and different behaviour as dimension grows.  Although a priori they require just as much tuning as \gls{Markov chain Monte Carlo}, the live points allow them to build proposal structures and apply clustering algorithms (analogously to ensemble samplers\cite{Goodman2010}), such that NS is naturally amenable to being robustly self-tuning. As they are guided by them, their reliability and efficiency usually improve when the number of live points is increased; see \ccite{Allison:2013npa,NSmethodsReview} for numerical investigations. For both region and step samplers, if no current live points lie inside a mode in a \gls{multi-modal} problem, that region of the constrained prior almost certainly won't be sampled, i.e., they miss that mode (see \cref{sec:choice_n_live}). 

Both samplers usually operate in the hypercube, a parameterization in which the prior is uniform over a unit hypercube (though this is not a requirement in NS). This slightly simplifies the problem to that of sampling uniformly from within a contour defined by the threshold. 
This is usually achieved by the inverse transformation method. In this case, users specify their \glspl{prior} by an inverse-cumulative density function rather than a density function. For example, suppose we desired a Gaussian prior for a parameter, $\param \sim \normaldist(\mu, \sigma^2)$. We could transform a unit hypercube parameter, $u \sim \uniformdist(0, 1)$, using the standard normal distribution's inverse-cumulative density function, $\Phi^{-1}$,
\begin{equation}
    \param = \mu + \Phi^{-1}(u) \, \sigma.
\end{equation}
This in principle allows all manner of \glspl{prior}, though see \nameref{sec:limitations} for further discussion.

\subsubsection{Region sampling}

In region sampling, we try to construct a region that bounds the \gls{iso-likelihood contour} defined by the threshold. To find the region, we construct a geometric shape around the current distribution of live points. The shape must contain at least the currently estimated volume. We then draw independent and identically distributed (iid) samples from within that region until we obtain a sample that passes the current \gls{likelihood} threshold. To be confident that the region did not encroach the contour, implementations of region sampling often expand the region by a user-specified factor or by a factor found through cross-validation. For example, dividing the live points into a training and test set and ensuring that the region found from the training set includes points in the test set~\cite{Buchner2014test,buchner2019bigdata}. The expansion factor improves reliability at the expense of efficiency.

The simplest region sampler would be sampling from the entire unit hypercube, that is the entire prior, which rapidly becomes prohibitively inefficient. Instead, most region samplers attempt to estimate the \gls{constrained prior} by wrapping the live points with one~\cite{Mukherjee_2006,Parkinson_2006} or more overlapping ellipsoids. Using more than one ellipsoid allows complicated and \gls{multi-modal} \glspl{iso-likelihood contour} to be efficiently bounded. An appropriate number of ellipsoids can be found by applying clustering algorithms to the live points to identify distinct \glspl{mode}, such as x-means (\ccite{Shaw_2007}, see \nameref{sec:cosmo}). The shape and location of the ellipsoids may be approximately found from the mean and covariance of the live points they contain, and their volumes may be judiciously expanded by a tuning parameter. This has successfully been implemented in \MN~\cite{MultiNest1}. Alternatively, \code{MLFriends}~\cite{Buchner2014test,buchner2019bigdata} places an ellipsoid around every live point, and determines the ellipsoid scale by bootstrapping (similar to kernel density estimation with a uniform kernel).

Region samplers suffer from two major limitations when the dimension or complexity of the contour grow. First, the ability to accurately bound a complicated contour strongly depends on the number of live points.
For instance, we may fail to identify a substructure with too few points, resulting in overly large regions that bound complicated substructures and in poor efficiency.
Alternately, since live points are distributed uniformly within the  \gls{constrained prior} rather than near the edge, wrapping a small number of live points can result in overly small estimates of the  \gls{constrained prior}.
Second, the accuracy and efficiency of region samplers suffer from the \gls{cod}.
To see how it strikes, consider region sampling with a single ellipsoid. Suppose that the true contour is a unit hyper-cube. The smallest ellipsoid that we could construct that enclosed the contour would be a sphere of diameter $\sqrt{\dimension}$. The volume of such a sphere blows up exponentially as dimension increases, leading to $\order{e^{-\dimension}}$ efficiency. This follows the general result that an exponentially increasing fraction of volume lies near the boundaries of high-dimensional shapes.
As a result, region samplers are efficient and practical only for moderate-to-low dimensionality ($\dimension \lesssim 20$).

\subsubsection{Step sampling}

Step samplers, by contrast, do not attempt to construct a region that bounds the \gls{iso-likelihood contour} and so avoid some of the issues described above. Instead, they evolve a randomly chosen existing live point through a sequence of steps to a new approximately independent position. The acceptance rule is simply that we accept a transition to $\params^\prime$ if
\begin{equation}
\like(\params^\prime) > \threshold
\end{equation}
i.e., each step must stay inside the contour (cf.~\cref{eq:mcmc}). Such step samplers are akin to running constrained \gls{Markov chain Monte Carlo} inside NS, and were in fact the originally proposed solution.
Strategies for generating new positions vary widely, and currently include:
\begin{itemize}
    \item random-walk Metropolis \cite{sivia2006data}, where new positions are proposed based on a local target distribution (e.g., a multivariate Gaussian),
    \item ensemble proposals \cite{Veitch2010}, which use the distribution of all live points to propose new positions using strategies such as differential evolution \cite{2006S&C....16..239T}
    \item slice sampling variants, where new positions within the \gls{constrained prior} are proposed along a randomly chosen principal axis (slice sampling; \cite{2012ASAJ..132.3251J,polychord}), or randomly chosen direction (hit-and-run; \cite{smith1984efficient,Zabinsky2013}), and
    \item gradient-based trajectories \cite{habeck2015nested,Betancourt2011,Skilling2012,nested_basin_sampling,2020MNRAS.493.3132S} that `reflect' off the current \gls{likelihood} constraint.
\end{itemize}
See \ccite{olander,NSmethodsReview,doi:10.1063/1.4985378} for further discussion. In step samplers with a step size parameter, such as random walk Metropolis, the step size is often tuned to ensure a substantial fraction ($>20\%$) of proposed positions are accepted. This avoids an unacceptable overall sampling efficiency \cite{sivia2006data,brooks2011handbook}.
This tuning may be performed continuously over the course of an NS run \cite{sivia2006data,polychord,2020MNRAS.493.3132S}, although this can introduce biases \cite{nested_basin_sampling}.

It can be challenging to judge the number of steps required to ensure that the live points are independent draws from the constrained prior (see \cref{sec:step_sampling} for further discussion). While mild violations of this requirement might be inconsequential \cite{2020MNRAS.493.3132S}, strong violations lead to unreliable NS \gls{evidence} estimates \cite{2019MNRAS.483.2044H}.
The number of iterations required for new samples to approximately de-correlate scales as $\order{\dimension^2}$ for random-walk proposals with tuned step sizes, and $\order{\dimension^{1} - \dimension^{5/4}}$ for slice sampling or gradient-based trajectories \cite{brooks2011handbook}.
In practice, the number of steps is often chosen to be  $a \dimension^{b}$, where $a$ is of order one and $b$ is the anticipated dimensional scaling.

The computational cost scales linearly with the number of steps. Thus unlike region samplers, step samplers escape the \gls{cod} as their cost shows only polynomial $\order{\dimension^b}$ scaling with dimensionality. Nevertheless, region samplers are often more efficient in low dimensions. As a result, step samplers are more often used when applying NS in high-dimensions ($\dimension \gtrsim 20$).

\Cref{tab:codes} compares the approaches in several publicly available NS implementations that originated in different research fields. For a multitude of programming languages, there are well-documented, free and open source codes. Support for parallelization to computing clusters and checkpointing are also common features. They usually work with the logarithms of the \gls{evidence} and \gls{likelihood}, as the latter may be numerically tiny such that it cannot be represented as a floating point value.  In general, sensible defaults for numerous region and step samplers have been found to work across a large variety of problems~\cite{2020MNRAS.493.3132S}. We present a simple numerical example in \cref{sec:example}.

\rowcolors{1}{white}{boxedtext}
\begin{table*}[thb!]
\centering
\begin{tabular}{l c c c c c}
\toprule
Code & Methods & Dynamic & Languages & Field & Pub.~Year\\%
\midrule
\urlcode{CosmoNest}{http://dparkins.github.io/CosmoNest/}~\citetable{Mukherjee_2006,Parkinson_2006} & ellipsoid & fixed & \code{Fortran} & Cosmology & 2006 \\
\MN~\citetable{MultiNest1,MultiNest2} %
& multi-ellipsoid & fixed & \makecell{\code{Fortran}, \\ \code{C/C++}, \urlcode{Python}{https://github.com/JohannesBuchner/PyMultiNest}
} & Cosmology & 2008 \\%
\urlcode{DIAMONDS}{https://github.com/EnricoCorsaro/DIAMONDS}~\citetable{2015EPJWC.10106019C} %
& multi-ellipsoid & fixed & \code{C++} & Astrophysics & 2015 \\
\urlcode{nestle}{https://github.com/kbarbary/nestle/}~\citetable{nestle} %
& ellipsoid, multi-ellipsoid & fixed & \code{Python} & Astrophysics & 2015 \\
\urlcode{nessai}{https://github.com/mj-will/nessai}~\citetable{nessai,Williams:2021qyt} %
& normalising flow ellipsoid & fixed & \code{Python} & Gravitational waves & 2021 \\
\midrule
\urlcode{(dy)}{https://dypolychord.readthedocs.io}\unskip\PC~\citetable{polychord,Higson2018} %
& slice & dynamic & \makecell{\code{Fortran}, \\ \code{C/C++}, \code{Python}} & Cosmology & 2015 \\
\urlcode{LALInferenceNest}{https://git.ligo.org/lscsoft/lalsuite/-/tree/master/lalinference}~\citetable{Veitch2015} & \makecell[{{cp{4cm}}}]{random walk, ensemble, \\ differential evolution} & fixed & \code{C} & Gravitational waves & 2015 \\
\urlcode{Nested\_fit}{https://github.com/martinit18/nested\_fit}~\citetable{Trassinelli:2016vej,proceedings2019033014,e22020185} %
& random walk & fixed & \code{Fortran} & Atomic physics & 2016\\
\urlcode{cpnest}{https://github.com/johnveitch/cpnest}\citetable{cpnest} %
& \makecell[{{cp{4cm}}}]{slice, differential evolution, \\ Gauss, Hamiltonian, ensemble} & fixed & \code{Python} & Gravitational waves & 2017 \\
\urlcode{pymatnest}{https://github.com/libAtoms/pymatnest}~\citetable{ConPresNS} %
& \makecell{random walk, Galilean, \\ symplectic Hamiltonian} & fixed & \code{Python} & Materials & 2017 \\
\urlcode{NNest}{https://github.com/adammoss/nnest}\citetable{Moss:2019fyi} %
& normalising flow random walk & fixed & \code{Python} & Cosmology & 2019 \\
\urlcode{DNest5}{https://github.com/eggplantbren/DNest5}~\citetable{brewer2011diffusive} %
& user-defined, random walk & diffusive & \code{C++} & Astrophysics & 2020 \\
\urlcode{BayesicFitting}{https://github.com/dokester/BayesicFitting}~\citetable{kester2021bayesicfitting} %
& random walk, slice, Galilean, Gibbs & fixed & \code{Python} &Astronomy & 2021\\
\midrule
\urlcode{dynesty}{https://github.com/joshspeagle/dynesty}~\citetable{2020MNRAS.493.3132S} %
& \makecell{ellipsoid, multi-ellipsoid, MLFriends \\ \& Gauss, slice, Hamiltonian} & dynamic & \code{Python} & Astrophysics & 2020\\
\urlcode{UltraNest}{https://github.com/JohannesBuchner/UltraNest/}~\citetable{2021JOSS....6.3001B} %
& \makecell[{{cp{4cm}}}]{MLFriends + ellipsoid \& Gauss,\\ hit-and-run, slice}  & reactive & \makecell{\code{Python}, \urlcode{Julia}{https://github.com/bat/UltraNest.jl}, \code{R},\\\code{C/C++}, \code{Fortran}} & Astrophysics & 2020\\
\urlcode{jaxns}{https://github.com/Joshuaalbert/jaxns}~\citetable{2020arXiv201215286A} %
& multi-ellipsoid \& slice & fixed & \code{jax} & Astronomy & 2021 \\
\bottomrule
\end{tabular}
\caption{\captiontitle{Comparison of NS codes} The first two groups are region samplers and step samplers, respectively, whereas the third group offers both.
Dynamic implementations allow the number of live points to be changed during a run. We show the language in which the NS code was written followed by any additional languages for which interfaces exist, and the field from which the code originated (though most are general purpose codes). 
}
\label{tab:codes}
\end{table*}

\subsection{Parallelization}\label{sec:parallel}

To utilize computing resources, we could parallelize the computation of the likelihood function. If that is impossible or impractical, we may wish to design an efficient scheme for parallelizing an NS run itself. 
As discussed in \ccite{higson2019dynamic,2020MNRAS.493.3132S}, statistically independent NS runs may be combined into an equivalent NS run with $\sum \nlive$ live points.
To achieve this, pool the initial live points from the independent runs $a$, $b$, $\ldots$ together, giving $\sum \nlive$ live points drawn from the prior. Remove the worst, supposing it lay at $\threshold$ and originated from run $a$. To draw a replacement from the prior subject to $\like > \threshold$, simply take the replacement used in run $a$, since it is already a draw from the prior subject to $\like > \threshold$. We may continue in this fashion, weaving together independent NS runs to build a new NS run with $\sum \nlive$ live points. This allows parallelization of an NS run with $\nlive$ live points into $\ncpu$ independent runs with about $\nlive / \ncpu$ live points each. The independent runs themselves proceed linearly and may be ultimately combined, resulting in a speedup of about $\ncpu$. Even simpler, estimates of the \gls{evidence} integrals may be combined by weighted averaging.

However, the reduction in the number of live points per run, $\nlive / \ncpu$, impacts the exploration schemes discussed in
\cref{sec:exploration},
especially for ellipsoidal rejection sampling where it could lead to inefficient or faulty bounding ellipsoids. It may therefore be desirable to utilize parallelization within individual NS runs. For example, by drawing $\ncpu$ candidate replacement points and evaluating their \glspl{likelihood} in parallel. We could subsequently replace the worst $\ncpu$ live points at each iteration~\cite{Burkoff2012,2014AIPC.1636..100H}, replace a single live point and consider the other evaluated points at subsequent iterations~\cite{polychord}, or replace a single live point and discard as many as $\ncpu - 1$ acceptable candidate points~\cite{MultiNest1}. The latter is wasteful if more than one viable point are likely to be found among the $\ncpu$ candidates,
\begin{equation}
    \text{Speed-up} \approx \min \left[\ncpu, 1  / \epsilon \right]
\end{equation}
If points are considered at subsequent iterations,
\begin{equation}
    \text{Speed-up} \approx \nlive \log \left(1 + \frac{\ncpu}{\nlive}\right)
\end{equation}
and so a speedup of about $\ncpu$ if $\ncpu \ll \nlive$. The expression originates from the fact that the threshold increases as the run progresses, meaning that points drawn from the  \gls{constrained prior} might not be valid at a subsequent iteration. Lastly, replacing $\ncpu$ points in parallel per iteration results in a speedup of about $\ncpu$ but increases the variance in the \gls{evidence} integral by about $\sqrt{\ncpu}$. Increasing $\nlive$ by a factor $\sqrt{\ncpu}$ to maintain the same uncertainty decreases the speedup to about $\sqrt{\ncpu}$.

\subsection{Stopping conditions}\label{sec:stopping}

We must decide when to stop an NS run. The fact that we can only perform a finite number of iterations introduces a truncation error in \cref{eq:sum_Z} that we wish to be negligible. Skilling originally proposed to stop NS once we reached the \gls{posterior} \gls{bulk} at $X \simeq e^{-H}$ at iteration $\niter \simeq \nlive H$, or using an estimate of the remaining \gls{evidence}. Popular NS software later adopted the latter. In \MN, this was based on the maximum \gls{likelihood} found so far, $\max \like X$, whereas \PC chose the mean \gls{likelihood}, $\bar \like X$. They stopped once $\Delta \ev / \ev \le \tol$, where $\tol$ is a user-specified parameter. Upon deciding to stop, the truncation error in the \gls{evidence} was corrected by either adding an estimate of the remaining \gls{evidence} or by killing the live points one by one without replacement and incrementing the \gls{evidence} in the usual manner until no live points remain. The latter is in keeping with the NS approach. For the former the remaining \gls{evidence} may be estimated by $\bar \like X$. See \ccite{2011MNRAS.414.1418K} for further discussion of the statistical properties of this estimate of the remainder.
When NS is used to calculate the \gls{partition function} of a material system (\cref{box:stat_mech}), physically motivated stopping conditions can be based on the expected minimum energy (negative log \gls{likelihood}) or the sampled temperature, which is proportional to the derivative of the limiting energy with respect to NS iteration~\cite{pt_phase_dias_ns}

None of these approaches guarantees that summation hasn't been terminated too early; there could well be a spike of enormous \gls{likelihood} lurking inward. The computational budget, furthermore, cannot be easily anticipated ahead of time. However, runs that are terminated prematurely may still be used to illustrate what was learned so far about the \gls{posterior}. Proposals to construct termination criteria for a fixed computational budget that result in unbiased estimates of the \gls{evidence} have been suggested~\cite{walter2015point}.

\section{Results}\label{sec:results}
NS results in an estimate of the integral in \cref{eq:Z} and, in the context of Bayesian statistics, a weighted set of draws from the \gls{posterior} distribution. We discuss some examples in \nameref{sec:applications} and how to report the results in \nameref{sec:repro}. The error estimate in \cref{eq:delta_Z} depends on the compression and cannot be known ahead of time. If the achieved error is unacceptable, the NS run can be repeated with more live points or combined with a new run. The shape of the \gls{posterior} can be found from the \gls{posterior} weights by density estimation. There are dedicated software packages for making publication quality figures of marginalized \gls{posterior} densities from weighted samples, including \anesthetic\cite{Handley:2019mfs}, \code{superplot}\cite{Fowlie:2016hew}, \code{pippi}\cite{Scott:2012qh}, \code{dynesty}\cite{2020MNRAS.493.3132S}, \code{getdist}\cite{Lewis:2019xzd}, \code{corner}\cite{corner} and \code{pygtc}\cite{Bocquet2016}.

There are a few ways to check the results. First, we can check the NS implementation, rather than the particular run. To do so, we can compute the \gls{evidence} integral in \cref{eq:Z} for problems with known analytic solutions; see \ccite{MultiNest2,Feroz:2013hea,beaujean2013initializing,polychord} for examples including a multi-dimensional Gaussian, an eggbox function, the Rosenbrock function\cite{10.1093/comjnl/3.3.175}, Gaussian shells and a mixture of a Gaussian and a log-gamma distribution.
Similarly, in some cases, the NS estimates of the volume variable at each iteration, $X(\threshold)$, may be checked against analytic results\cite{Buchner2014test}. If discrepancies are found, the implementation is suspect. Alternatively, we can repeat calculations and check whether the distribution of results is consistent with what would be expected if the software was working correctly. \Ccite{2019MNRAS.483.2044H} describes procedures for doing this, including tests requiring only two NS runs; these are implemented in \urlcode{nestcheck}{https://github.com/ejhigson/nestcheck}~\cite{nestcheck}. 

Second, the particular NS run of interest may also be checked using a test of the \gls{insertion indexes} of new live points\cite{Fowlie:2020mzs}. If NS draws new live points independently from the  \gls{constrained prior}, as it should, the ranks in \gls{likelihood} of each new live point compared to the current live points should be uniformly distributed. This test is implemented in the \anesthetic~\cite{Handley:2019mfs} NS analysis software, which is compatible with \PC and \MN, and used on the fly in the \urlcode{nessai}{https://github.com/mj-will/nessai}~\cite{nessai,Williams:2021qyt} and \urlcode{UltraNest}{https://github.com/JohannesBuchner/UltraNest/}~\cite{2021JOSS....6.3001B} NS implementations. If this test fails, but other implementation checks pass, the choices of exploration strategy for the problem at hand may be inadequate. Similarly, \ccite{stokes2016,stokes_thesis} discuss testing whether live points are uniformly distributed in the unit hypercube in two-dimensional problems. Last, in the context of parameter inference we can compare the posterior samples obtained from different NS implementations or from \gls{Markov chain Monte Carlo} and NS (see for example \ccite{romeroshaw2020}) or perform \gls{sbc} to check expected properties of the posterior.

As discussed in \nameref{sec:stopping}, most NS implementations stop once the estimate of the remaining \gls{evidence} appears negligible. This could omit spikes in \gls{likelihood} lying inside the remaining unexplored volume. If that is a concern, it may be beneficial to optimize the \gls{likelihood} using a local optimization algorithm starting from the live point with the greatest \gls{likelihood}. The run may be restarted with stricter stopping conditions if local optimization finds a maximum \gls{likelihood} that is orders of magnitude greater than that found during the NS run.

Lastly, in many settings we may be concerned about the possibility that \glspl{mode} were missed. While the tests described above may indicate whether implementation errors are present, it is in general impossible to know for certain whether all \glspl{mode} were identified. We can inspect the \gls{posterior} by eye, or perform a \gls{mode} identification algorithm on the \gls{posterior} samples. If the expected number of \glspl{mode} was known and \glspl{mode} are evidently missing, the number of live points should be increased.

\section{Applications}\label{sec:applications}

Here we present the most established NS applications, highlighting its advantages in each case. Besides these established applications, NS is beginning to be utilized in many other areas, including signal processing\cite{combined_NS}, phylogenetics\cite{phylogenetics_ns}, systems biology \cite{biology_ns,ns_systems_bio}, acoustics\cite{Beaton2017,VanSoom2020}, nuclear physics\cite{Lewis2014,FRS-ESR:2019pha}, atomic physics\cite{Trassinelli:2016kki,Trassinelli:2016vej,Covita:2017gxr,DeAndaVilla2019,Machado:2018cxl}, exoplanet searches\cite{2015MNRAS.448.3206B,2017AJ....154...91L,2018MNRAS.479.2968H,2020ApJ...890..174K,2021MNRAS.503.1248A}, and geology\cite{Elsheikh2013}.

\subsection{Cosmology}\label{sec:cosmo}

The rapid spread of Bayesian methods in cosmology~\cite{Trotta:2008qt} in the early 2000s was generated by the growth of data, specifically the new cosmic microwave background (CMB) temperature power spectrum measured by the Wilkinson Microwave Anisotropy Probe (WMAP) \cite{Spergel_2003}, and the type-Ia supernovae distance measurements \cite{Riess_1998, Perlmutter_1999}. However,  these new and powerful cosmological datasets had raised difficult questions regarding the cosmological model. The favoured model of the Universe included a mysterious accelerating force, the dark energy, which accounted for 70\% of the energy density today. And the initial spectrum of density fluctuations in the early Universe, which were shown to be Gaussian and adiabatic by WMAP, indicated a period of accelerated expansion at very early times (about $10^{-23}$ seconds after the Big Bang), known as cosmic inflation \cite{PhysRevD.23.347}. Finally, there was still the question of the missing mass of the Universe, which generates the gravitational fields required for cosmic structure, known as cold dark matter (CDM). All three of these phenomena (cold dark matter, dark energy, and cosmic inflation) had proposed explanations from the field of high-energy theoretical physics, and these model predictions could be combined with the new wealth of cosmological data to be evaluated with respect to each other, model by model. Thus model selection in general, and the computation of the Bayesian \gls{evidence} using NS specifically, became a tool of choice.

The simplest model of cosmic inflation is one driven by a single scalar field, a particle physics object similar (but not identical) to the Higgs boson. The behaviour of the scalar field driving inflation is determined by its potential $V(\phi)$ ($\phi$ being the value of the scalar field)\cite{PhysRevD.23.347,Liddle_1994}. The formulation of $V(\phi)$ (as a function of some set of parameters $\Theta$) determines the duration of inflation along with the distribution of anisotropies found in the CMB. \Ccite{Martin_2011} performed a detailed \gls{Bayesian model comparison} between 193 inflationary models using the region sampler \MN. They used the cosmological observations (including CMB data from WMAP mission) to discriminate between alternative models for $V(\phi)$ and found slight preference for so-called Small Field Inflation (SFI) models over Large Field Inflation (LFI) models.

\begin{figure*}
  \centering
  \includegraphics[width=\textwidth]{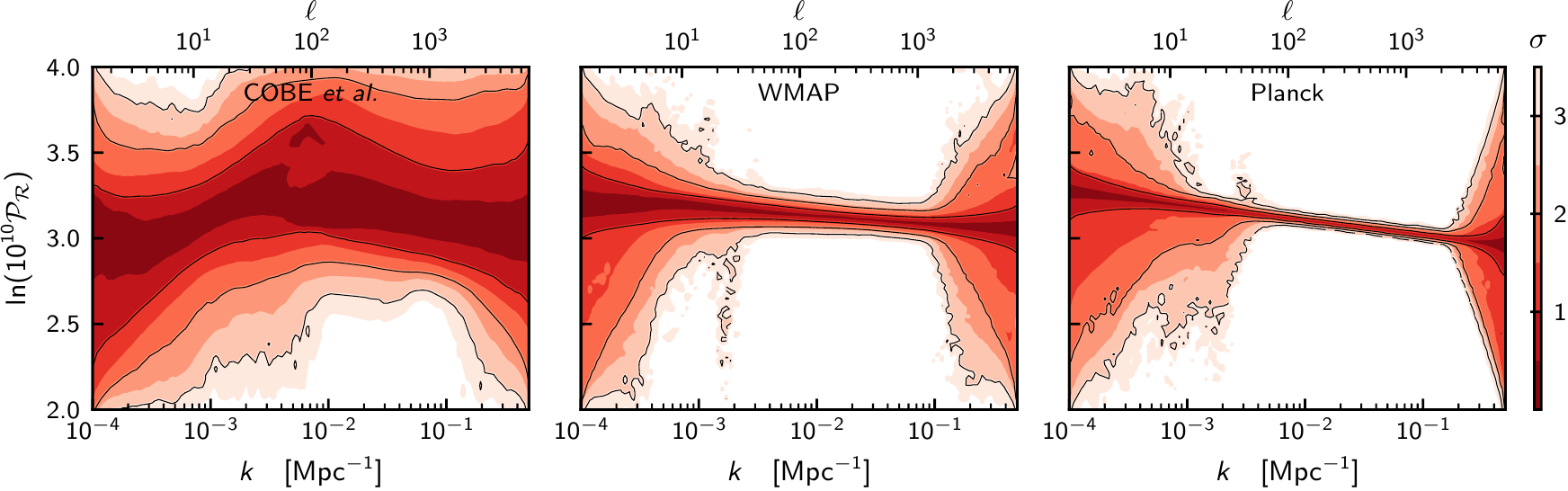}

  \vspace{10pt}

  \includegraphics[height=0.27\textwidth]{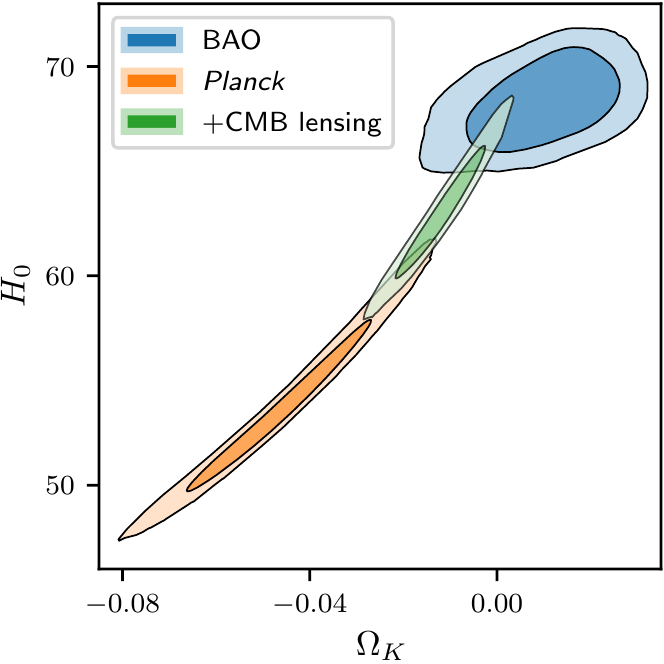}
  \includegraphics[height=0.27\textwidth]{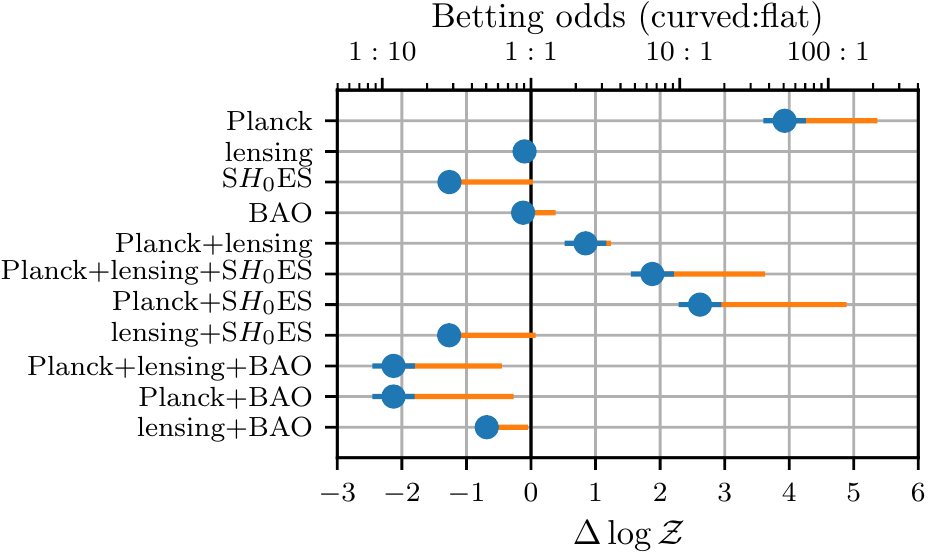}
  \includegraphics[height=0.27\textwidth]{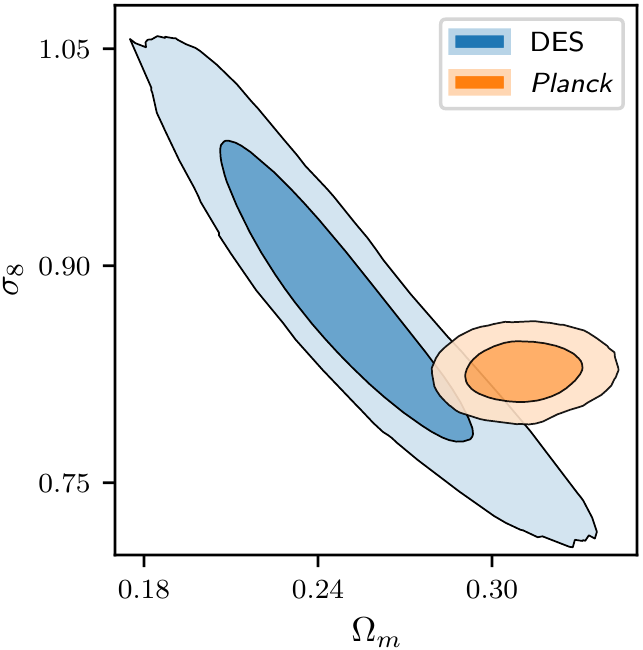}
  \caption{\captiontitle{Cosmological applications of NS} \emph{Top row:} Cosmological non-parametric reconstruction of the power spectrum of primordial cosmological fluctuations, as measured by cosmic microwave background satellites across human history. The reconstruction uses a linear spline-based procedure with $N$ movable knots. Each of the locations of the knots along with the cosmological and nuisance parameters are varied in a full NS fit. The \gls{evidence} is then used to marginalize over $N$ to produce the final plots. The standard model of cosmology predicts a featureless tilted power spectrum, so such non-parametric reconstructions are of great interest for astrophysicists searching for \gls{evidence} of beyond standard model physics.
  \emph{Bottom row:} NS in cosmological tension quantification. Bayesian \gls{evidence}s computed by NS can be used to quantify the level of disagreement which may be hidden by marginalization of high-dimensional parameter spaces. \emph{Bottom left:} Planck CMB data is in tension with CMB lensing and BAO in the context of curved cosmologies, so caution should be exercised in combining them~\cite{2019arXiv190809139H}. If one includes only CMB data, a \gls{Bayesian model comparison} shows preference for a closed Universe relative to a flat one (\emph{Bottom middle}) in spite of the Occam penalty associated with the additional parameter (orange bar). \emph{Bottom right:} There is also tension between weak lensing data (DES) and the CMB (Planck). }
  
\label{fig:cosmo}
\end{figure*}

Another area where NS has been used extensively in cosmology is the modelling of galaxy clusters \cite{Allen_2002,Allen_2011}. Clusters of galaxies are the most massive gravitationally bound objects in the Universe and therefore, can be used to trace the formation of large scale structure in the Universe. Galaxy clusters can be observed through several methods including X-ray observations, weak gravitational lensing and by exploiting the \gls{SZ}. Weak gravitational lensing involves the distortion of the images of the background galaxies by the presence of a large mass lying along the line-of-sight. Weak lensing allows to probe the total mass distribution, including the dark matter, of the galaxy cluster. 
\Ccite{Feroz_2009_SZ} presented a Bayesian approach using NS for the detection of galaxy clusters in \gls{SZ} data, including estimation of parameters associated with the physical model assumed for the cluster and quantification of the detection using Bayesian model selection, making use of the statistics of the CMB anisotropies in the \gls{likelihood} function. \Ccite{Hurley_2011} presented a joint Bayesian analysis of weak lensing and \gls{SZ} observations of several galaxy clusters, allowing for the estimation of gas fraction of individual clusters.
In modern weak lensing surveys~\cite{2020A&A...638L...1J} such as DES~\cite{2018PhRvD..98d3526A} and KiDS~\cite{2021A&A...645A.104A}, NS forms a critical part of their parameter estimation, model comparison and tension quantification pipelines~\cite{2019PhRvD.100d3504H} (\cref{fig:cosmo}).

The expansion history of the Universe can also be measured through the use of distance measurements, such as observations of type Ia supernovae (SNIa), which can be used as `\gls{standard candles}'. The important results from the Supernova Cosmology Project and the High-Z Supernova Search Team \cite{Riess_1998, Perlmutter_1999} presented \gls{evidence} for the accelerated expansion of the Universe, requiring the existence of a mysterious dark energy acting against gravity. 
Estimation of cosmological parameters from the observations of SNIa light curves has historically been done using a $\chi^2$ approach (see e.g.~\cite{Conley_2010}) which lacks a rigorous approach for the determination of systematic uncertainties. \Ccite{March_2011} introduced a Bayesian hierarchical model for the determination of cosmological parameters.
By using NS, they showed that their principled Bayesian approach delivered tighter statistical constraints on the cosmological parameters over 90\% of the time, reduced statistical bias by a factor $\sim$ 2-3 times and that it had better coverage properties than the standard $\chi^2$ approach.

The measurement of the CMB anisotropies by Planck~\cite{2014A&A...571A...1P} was an increase in the statistical power over the previous experiment (WMAP), but required more sophisticated modeling of galactic foregrounds and instrumental calibration, introducing $\order{20}$ ``nuisance parameters''. 
Metropolis-Hastings techniques for parameter estimation were able to accommodate these by exploiting the fact that these parameters were ``fast'' in comparison to the cosmological ones; namely, by caching results from previous calculations, these parameters can be changed with negligible computational cost providing the cosmological parameters remain fixed. 

Region-based samplers such as \MN are not able to exploit this hierarchy of parameter speeds, and could not navigate the now high-dimensional cosmological + nuisance parameter space reliably. To address this, cosmologists turned to step sampling based strategies as instantiated in the \PC algorithm~\cite{polychordcosmo,polychord} which uses slice sampling.\footnote{Axial slice sampling had been applied in the then near past to systems biology~\cite{aitken2013nested} although unsurprisingly the cosmological authors were not aware of this work. \PC improved upon the existing slice-sampling method by implementing covariance-based non-axial steps and \gls{mode} clustering in addition to the ability to exploit a hierarchy of parameter speeds.} 
These were successfully applied throughout the 2015 Planck inflation paper~\cite{2016A&A...594A..20P} to non-parametric reconstructions~\cite{2019PhRvD.100j3511H} (\cref{fig:cosmo}) and general inflationary model comparison, in particular to the challenging example of axion monodromy models. The non-parametric reconstruction approach, which only became possible with the ability to reliably (and fully) explore the parameter space of a large number of parameters, was an important model-independent demonstration of the simple power-law behavior of the primordial power spectrum of density perturbations.

Although step sampling was originally introduced in cosmology to exploit a fast-slow hierarchy of parameters, the better dimensional scaling opened up a new range of cosmological analysis possibilities that were inaccessible to region samplers. It has been applied to constraining kinetically dominated inflation models~\cite{2019PhRvD.100b3501H}, model comparison for the quantum mechanical initial conditions for the Universe~\cite{2021arXiv210403016G}, additional reconstructions of the dark energy equation of state~\cite{2017NatAs...1..627Z,2017MNRAS.466..369H}, astronomical sparse reconstruction~\cite{2019MNRAS.483.4828H}, and played a critical role in the GAMBIT combined cosmology and particle analyses~\cite{2021JCAP...02..022R,2020arXiv200903287G}. Step sampling takes a leading role in the REACH 21cm global cosmology analysis~\cite{2020arXiv201009644A}, and at the other end of the astrophysical scale it has also been applied to high-dimensional exoplanet analyses~\cite{2018MNRAS.479.2968H,2021MNRAS.503.1248A}.

\subsection{Particle Physics}
Particle physics is a field related to cosmology that has also seen various applications of NS.
Around 2010, when NS tools such as \MN were reaching maturity, the particle physics community was focused on the first results from the Large Hadron Collider (LHC). A particularly favoured theoretical framework was supersymmetry (SUSY; see e.g., \ccite{Martin:1997ns}). SUSY introduces an array of new particles and unknown parameters that can be fitted to collider data as well as observations from other experiments in a so-called global fit. The physics goals of a global fit are to understand what the model predicts in future experiments and how best to discover the new particles.

These global fits of \order{10} free parameters in SUSY models presented a problem for which NS tools were naturally well-suited, as they were typically \gls{multi-modal}.  A package named \code{SuperBayeS}~\cite{deAustri:2006jwj,Trotta:2008bp,Feroz:2011bj} utilized the \MN implementation of NS and was used to make a number of early LHC predictions and fits\cite{Trotta:2010mx,AbdusSalam:2009qd,Strege:2011pk,Buchmueller:2013rsa,Fowlie:2012im,Fowlie:2013oua}. As the LHC results have poured in throughout the 2010s, the theoretical landscape shifted and a wider set of models have been considered using NS (see e.g., \ccite{Catena:2014uqa,deVries:2015hva,Hernandez:2016kel,Kreisch:2019yzn}).
More recently, the \code{GAMBIT} collaboration has driven many such global fits making use of \MN and \PC to sample parameter spaces and compute Bayesian \gls{evidence}s, and benchmarked NS against a number of other MC based and gradient-free methods~\cite{scannerbit,Balazs:2021uhg}.
Related problems such as tuning phenomenological parameters in event generators have seen preliminary work, and could be rich avenues of NS application in the field.
Lastly, we note that NS was recently applied to the sampling space, rather than the parameter space, of a statistical model. This enables efficient computation of small $p$-values that are used in the discovery of new particles at the LHC\cite{Fowlie:2021gmr}.

\subsection{Gravitational waves}\label{sec:gw}
Gravitational-wave (GW) astronomy is a field that has blossomed since the first observation in 2015 of two colliding black holes~\cite{GW150914} by the LIGO \cite{2015CQGra..32g4001L} and Virgo \cite{2015CQGra..32b4001A} interferometers.
The signals are produced by non-axisymmetric changes in the gravitational field, typically sourced by the rapid motion of neutron stars and black holes in binary systems \cite{LIGOScientific:2021djp}.
The binary orbits decay through GW emission, increasing the orbital frequency and the rate of energy loss until the objects merge.
Such events, known as compact binary coalescences (CBCs), produce signals at frequencies of 10--1000\,Hz which are recorded in the detectors as a time-series.

Early development of Bayesian methods for GW analysis was contemporaneous with the first publication of NS, and it was found that NS provided an efficient means to robustly sample the \gls{posterior} distributions of GW signals~\cite{Veitch:2008ur}.
This is important as the \glspl{posterior} provide a rich new astrophysical view: from measuring the masses of neutron stars to the expansion rate of the Universe itself.
The small signal-to-noise ratios of observed signals
and degeneracies in the model parameter space produce \gls{posterior} distributions that are often \gls{multi-modal} and highly correlated.
To date, only NS and \gls{Markov chain Monte Carlo} approaches have been able to robustly sample the \glspl{posterior} of all observed events.
While each has its own advantages, agreement between the NS and \gls{Markov chain Monte Carlo} approaches has been critical in building confidence in results.
However, the NS approach is better-suited to providing robust \gls{evidence} estimates for model comparisons \citep{LIGOScientific:2019eut}; the efficiency of the NS approach does not depend so much on the use of problem-specific proposals; and massively parallel approaches using the \code{dynesty} code have made analyses with more advanced signal models \citep{pbilby} computationally tractable.
The success of NS in analysing merging binary signals has inspired many other efforts
such as the analysis of continuous signals from individual rapidly-rotating
non-axisymmetric neutron stars \citep{2017arXiv170508978P, 2019ApJ...879...10A};
detection of unmodelled sources \cite{2017PhRvD..95j4046L}; model selection between
different physical mechanisms of core-collapse supernova encoded in the GW signal
\cite{2016PhRvD..94l3012P}; and detection of a stochastic superposition of weak merger
sources \cite{PhysRevX.8.021019}.

CBCs consisting of two black holes have been the dominant sources of GW signals seen
by the LIGO and Virgo detectors \cite{LIGOScientific:2021djp}.
The signals are modelled through a combination of post-Newtonian approximations to General Relativity and relativistic numerical modelling \cite{2007LRR....10....2F, 2014LRR....17....2B, 2014GReGr..46.1767H, 2016LRR....19....2B}.
The signal model, as observed in a detector, is parameterized by 8 parameters that are intrinsic to the binary system (the individual masses, $m_1$ and $m_2$, and their 3D spin vectors) and 7 parameters related to the relative orientation and position of the system with respect to the detector (including the source's luminosity distance $D_{\rm L}$ and location on the sky).
For systems that include at least one neutron star, the signal model's phase evolution requires additional parameters related to the neutron star equation-of-state \cite{BAIOTTI2019103714}.

\begin{figure}[thb!]
\begin{subfigure}{\linewidth}
    \centering
    \includegraphics[width=0.8\linewidth]{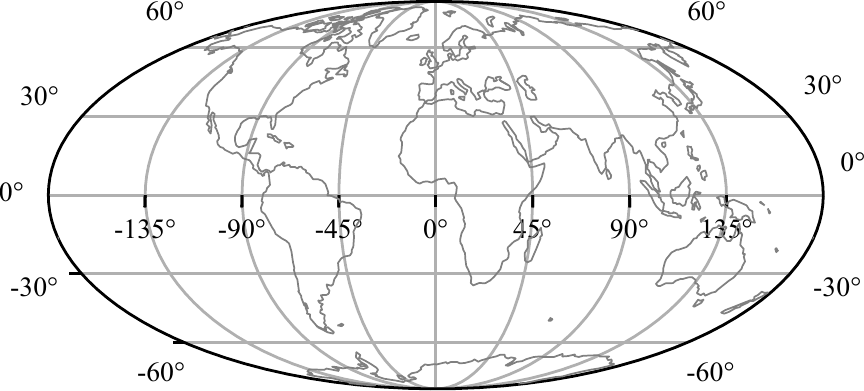}
    \vspace{0.25cm}
    \subcaption{\captiontitle{Sky location of the source of the binary black hole merger signal GW151226} The \gls{posterior} probability distribution from NS is shown in orange\cite{2016PhRvL.116x1103A}.}
    \label{fig:gwparameters}
    \includegraphics[width=0.9\linewidth]{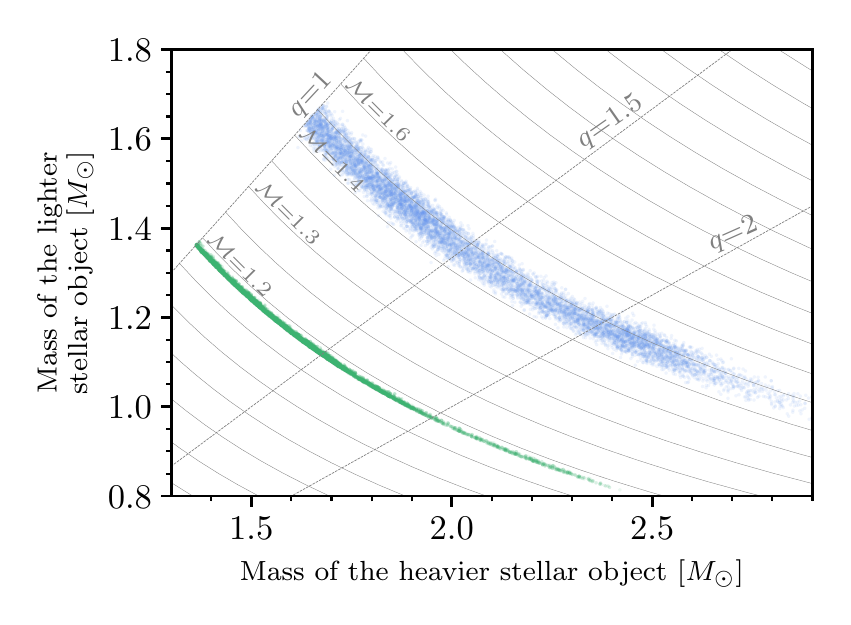}
    \subcaption{\captiontitle{Masses of stellar objects involved in two binary neutron star mergers} The \gls{posterior} samples from NS are shown for the neutron star mergers GW170817 (green) and GW190524 (blue). Solid grey lines mark curves of constant chirp mass $\mathcal{M}$, while dashed lines mark curves of constant mass ratio $q$. When viewed in $\mathcal{M}$-$q$ space, the posteriors show only weak correlation.
    }
    \label{fig:gw_masses}
    \end{subfigure}
    \label{fig:gws}
    \caption{\captiontitle{Applications of NS in GW astronomy}}
\end{figure}

Signals from black hole mergers last only a few seconds within the sensitive frequency regime of the detectors. On these timescales the detector data can be approximated as having noise drawn from a stationary Gaussian process described by a known power spectral density, which can either be estimated from data surrounding the signal \cite{Veitch2015}, or inferred directly using a parameterized model of its shape \cite{2021PhRvD.103d4006C}. For inference, a Gaussian \gls{likelihood} function of the form given by Whittle \cite{whittle1957} can be used \cite{Veitch2015}. For multiple detectors, which have independent noise, the \glspl{likelihood} can be coherently combined using the product rule of probability. The prior distributions used are discussed in \ccite{Veitch2015}, but are generally set to be uninformative (e.g., uniform over a sphere for sky coordinates), or constrained to be within a physically reasonable range.
In addition to the parameters related to the source, the \gls{likelihood} can 
also contain $\sim$ 60 unknown parameters that relate the 
frequency-dependent uncertainties in the phase and amplitude calibration 
of the detectors \cite{2012PhRvD..85f4034V}.

For an observed signal, using the \gls{likelihood} and \glspl{prior} discussed above, the joint \gls{posterior} of all these 15 (or more) parameters can be extracted after application of NS.
The \gls{posterior} distribution of the typical events observed so far are not \gls{uni-modal} or Gaussian.
Due to the relatively small signal-to-noise ratio, they exhibit significant degeneracies and correlations.
For example, the \gls{posterior} of the source sky location is largely determined by differences in arrival time of the signal at different detectors.
This produces ring-shaped degeneracies as demonstrated in \cref{fig:gwparameters}.


The many degeneracies encountered for typical CBC inference have led to significant work to identify optimal parameterization and, where possible, utilized a marginalized \gls{likelihood} (see \ccite{thrane2019} for a review).
An optimal parameterization involves identifying a mapping between the physical model parameters and combinations of these which reduce the complexity of the target density.
Usually, a good reparameterization involves identifying the combinations of physical parameters which are `best measured'.
As an example, the physical mass of the two component stellar objects are $m_1$ and $m_2$.
However, there is a strong banana-like correlation between the two masses (see, e.g.\ \cref{fig:gw_masses}) and an exact degeneracy under exchange of $m_1$ and $m_2$.
To enable efficient sampling, Veitch \textit{et al} \cite{Veitch2015} proposed sampling in the \emph{chirp-mass}, $\mathcal{M}$, and \emph{mass ratio} $q$: two algebraic combinations of the component masses.
In  \cref{fig:gw_masses}, we show that the \gls{posterior}, as viewed in $m_1-m_2$ space, follows contours of the chirp mass.
Physically, this is because the chirp-mass is the most well measured mass parameter (i.e.\ has the smallest \gls{posterior} width) followed by the mass ratio.

The chirp-mass and mass ratio parameterization, along with numerous others have greatly improved sampling efficiencies.
We note that the choice of sampling parameters, which we choose for computational efficiency, is separate from the choice of prior distributions.
If required, a Jacobian transformation \cite{Callister:2021} can be made to enable sampling in the optimal sampling parameters while setting \glspl{prior} on the physical parameters.

\subsection{Materials science}\label{sec:materials}

As hinted at in \cref{box:stat_mech} above,
NS can be used to study the thermodynamic properties of molecules and materials, which ultimately derive from the \gls{partition function}
\begin{equation}
Z(\beta) = \int e^{-\beta H(q, p)} \, \diff q \,\diff p,
\end{equation}
where \( q \in \mathbb{R}^{3N}\) specifies the spatial coordinates of the \(N\) particles in the system,
\(p \in \mathbb{R}^{3N}\) is the corresponding momentum vector,
\(H(q, p)\) is the Hamiltonian,
and \(\beta=1/k_BT\) is the inverse temperature.

The classical Hamiltonian can be separated into the configuration dependent potential energy, \(U(q)\), and momentum dependent kinetic energy, \(K(p) = \sum p^2 / 2m\) (where $m$ is the mass of each particle), giving
\begin{equation}
    Z(\beta) = \int e^{-\beta K(p)} \,\diff p \int e^{-\beta U(q)} \,\diff q.
\end{equation}
(For the fully quantum treatment using NS, see \ccite{quantum_part_func}.)
The first factor is a Gaussian that may be computed analytically,
\begin{equation}
Z_p  \equiv \int e^{-\beta K(p)} \,\diff p =  \frac{1}{N!} \left( \frac{2\pi m}{\beta \hbar^2}\right)^{3N/2},
\end{equation}
and we tackle the second, configuration dependent factor by NS. From the \gls{partition function} all thermodynamic quantities of relevance can be calculated, for example, the average energy,
\begin{equation}
    \left\langle H(q, p) \right\rangle = -\frac{\partial \ln Z(\beta)}{\partial \beta}
\end{equation}
and the heat capacity
\begin{equation}
    C_V = \frac{\partial \left\langle H(q, p) \right\rangle}{\partial T},
\end{equation}
and these may be interpreted as \gls{posterior} expectations to be computed after the NS run. In this application,
\begin{itemize}
    \item the factor $e^{-\beta U(q)}$ plays the role of the \gls{likelihood} function, parameterized by the inverse temperature, \(\beta\);
    \item the prior, according to the ergodic hypothesis, is uniform;
    \item the dimensionality is commonly on the order of $10^2$ to $10^3$ or more.
\end{itemize}
The striking difference from standard Bayesian inference is the inverse temperature parameter \(\beta\) in the \gls{likelihood}. Rather than having a single inference problem for some fixed value of \(\beta\), almost always we want to know the behaviour of observables {\em as a function of temperature}, so effectively we have a continuous family of inference problems. And for this, NS (and other density-of-states methods, see \cref{box:stat_mech}) have a remarkable feature: the \gls{likelihood} is a monotonic function of \(\beta\) and hence the entire NS algorithm is invariant to changes in \(\beta\). In practice this means that a single NS run can be used to calculate observables at all temperatures. 
For large $\beta$ values (corresponding to low temperatures), the \gls{partition function} is dominated by the lowest energy minima (highest \gls{likelihood} \glspl{mode}), and for molecules and materials these configurations correspond to the globally stable structures.
On the other hand, when \(\beta\) is small (corresponding to high temperatures), the \gls{partition function} is dominated by the large volume associated with high energy states --- or in the parlance of materials science, ``entropic effects''\cite{SciortinoKT00,BasinSampling}.

Some of the most interesting phenomena in molecular and materials science (and more generally in statistical mechanics) are associated with the above change of regime.  Collectively known as {\em phase transitions}, they are characterized by a dramatic change of where the \gls{bulk} of the \gls{posterior} mass lies, as the temperature (or other system parameters such as pressure) is varied --- and this is what makes it very challenging to study them using numerical sampling schemes. Experimentally, phase transitions are often observed indirectly as changes in the expectation value of observables (e.g.\ discontinuously, in the case of first order phase transitions such as melting and evaporation), or more directly as sharp peaks in response functions, such as the heat capacity or magnetic susceptibility. The promise of NS is to enable the calculation of such response functions in general with high reliability and the minimum of fuss.

There are some aspects of the materials application of NS that turn out to be favourable in comparison with the general inference problem. The first is a rather easy stopping criterion for the NS iterations. For any given model of the potential energy, it is generally not hard to come up with a good global lower bound on the energy, which translates into an upper bound in the \gls{likelihood}. As the \gls{likelihood} values sampled by NS appear to converge, if this is close to the known bound, the iterations can be stopped without the risk of missing the highest \gls{likelihood} mode.

Secondly, convergence of NS with the number of live points and other sampling parameters is desired and observed in terms of convergence of the heat capacity peak locations, and this typically occurs far earlier than the decorrelation of the sampler chains that are used to explore the  \gls{constrained prior}.

\subsubsection{Thermodynamics of LJ Clusters}

The Lennard-Jones (LJ) potential is a simple model for describing the pairwise interactions between atoms and it provides the basis for benchmarking algorithms for modelling materials. In particular, specific sized clusters of LJ atoms exhibit complex thermodynamic properties due to solid-solid transitions
(the finite system analogue of first order phase transitions mentioned above) caused by existence of competing low energy minima,
in addition to solid-fluid melting \cite{Wales03,WalesB06,BasinSampling}.

\begin{figure}[!ht]
\begin{subfigure}{\linewidth}
\centering
\includegraphics[width=7cm]{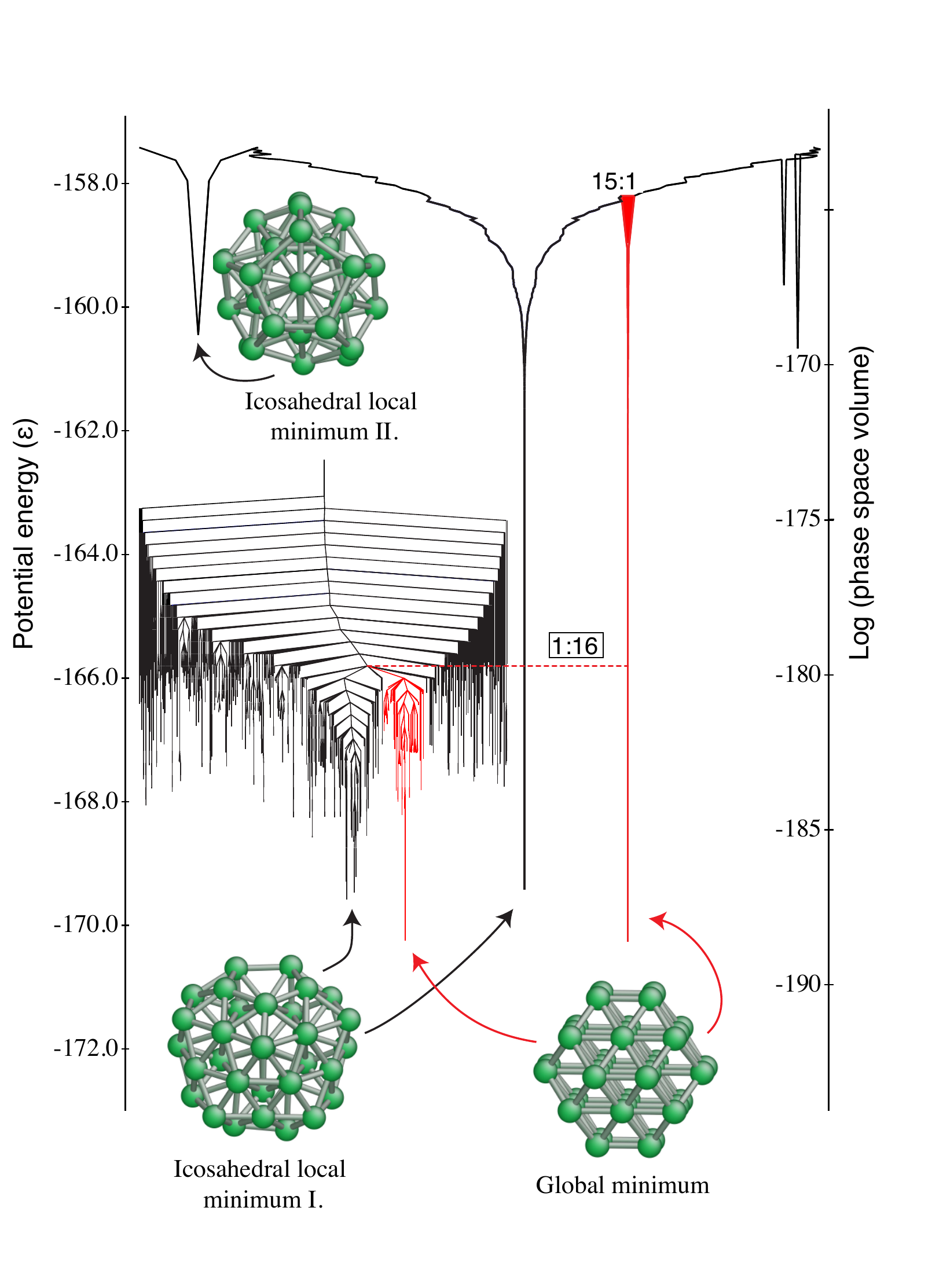}
\subcaption{
Potential energy landscape of LJ$_{38}$, showing the energy landscape chart (see text) obtained by NS, and the disconnectivity graph calculated using discrete path sampling, showing the 200 lowest energy minima\cite{LJ38_disconnectivity}. The global minimum basin is shown in red, and the relative phase space volumes of the basins at the separation point and the lowest known connecting path are indicated by the ratios.}
\label{fig:LJ38_PES}
\vspace{1cm}
\includegraphics[width=8.5cm]{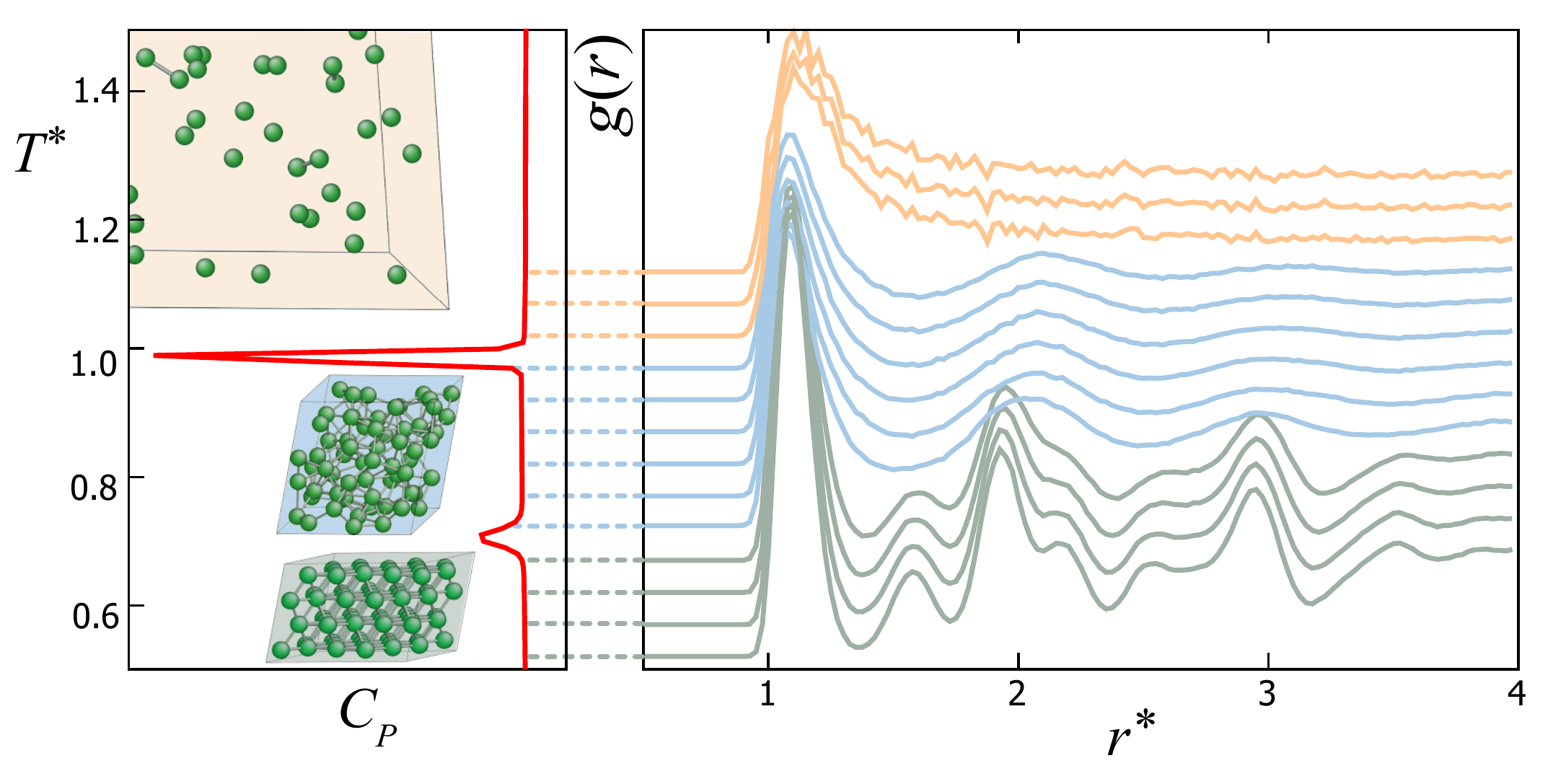}
\subcaption{
Constant pressure heat capacity curve (red) peaks are used to locate  transitions between the gas (top), liquid (middle) and solid (bottom) phases. Sampled configurations and weights enable calculations of temperature dependent observables such as the radial distribution function (right panel) \cite{pt_phase_dias_ns} (note that starred variables correspond to quantities in reduced units).}
\label{fig:PhaseDiagram}
\end{subfigure}
\caption{\captiontitle{Illustrations of materials science applications}}
\label{fig:materials}
\end{figure}

We illustrate some features of atomistic energy landscapes in \cref{fig:LJ38_PES} using the example of the cluster of 38 particles (\LJ{38}) along with the corresponding {\em disconnectivity graph}~\cite{BeckerK97,walesmw98} which help to convey the relationships between the large number of local energy minima (likelihood \glspl{mode}). Each leaf of the tree structure corresponds to a local minimum of the energy (or to a closely related set of them), and junctions represent saddle points connecting the minima. As the number of particles in the system is increased, the number of distinct local minima grows exponentially. According to one estimate, \LJ{31} for example has
\(10^{15}\) distinct minima, not counting permutation isomers.\cite{BasinSampling}

Potential energy disconnectivity graphs are good at showing the topology of the energy minima, but to reflect the volumes' free energy, disconnectivity graphs are required~\cite{KrivovK02,EvansW03}. Using NS, {\em energy landscape charts}\cite{partay2010efficient} can be generated, and we show one for \LJ{38} in \cref{fig:LJ38_PES}. The vertical axis is the potential energy, and the curve shows a series of (possibly nested) basins, whose width at each potential energy level is proportional to the volume of the corresponding \gls{posterior} \gls{mode} slice. In order to fit the chart in one diagram, the volumes are also scaled by an exponential factor whose logarithm is shown on the right hand axis. Each basin, encompassing a range of closely related configurations, corresponds to a macroscopic ``state'' of the system --- illustrative structures are shown.
The relative volume of the funnel associated with the global minimum (shown in red) and the entropically dominant minimum (in the centre) is \(1:15\) at the energy level where the two separate (at the resolution of this particular NS run), whereas the relative volume reverses to \(16:1\) at the energy level where the known lowest barrier path actually connects the two states\cite{lis87,walesd97a,Waless99}. 

It is through these landscape charts that the challenges of thermodynamic sampling can be best understood. If there are not enough live points, even though a few may make it into the global minimum basin where that splits off, there is a danger of that population of walkers ``dying out'' much before the global minimum is reached. This extinction can happen even while the different basins are still connected in principle, but the volume of states that connect them is so small (or rather, the path connecting them is so narrow) that in practice there is no communication between the basins. This phenomenon is generally referred to as \emph{broken ergodicity} \cite{BasinSampling}.

Note that while small LJ clusters serve as illustration and allow one to develop, benchmark and parameterize various details of sampling algorithms, NS is not the most efficient method for calculating observables in this case. Specialized ``bottom-up'' algorithms such as Basin-Sampling\cite{BasinSampling} are significantly more efficient, because they start from a low lying minimum (not necessarily the global minimum) and build a database of all the neighbouring minima by a series of  perturbations of the particles and subsequent relaxation. The performance of NS for LJ clusters can be somewhat improved if either (i) the sampler is allowed the use of the pre-generated database of local minima, as in the superposition-enhanced NS (SENS)\cite{superposition_enhanced}, or (ii) a large number of independent NS runs each with a single live point are suitably combined, as in Nested Basin-Sampling\cite{nested_basin_sampling}, but neither of these enhancements makes NS competitive for studying small and moderate sized particle clusters.

There is one case for which NS appears to be an effective tool for studying even the smallest clusters, and this is the determination of thermodynamically favourable \emph{transformation paths} between different states at moderately high temperatures where harmonic transition state theory is no longer applicable.\cite{nested_paths}

\subsubsection{Phase diagrams of materials}
\label{sec:phase_diagrams_of_materials}

It is in the study of condensed phase systems that NS really comes into its own. There are two reasons for this: one is that it is generally difficult to make efficient sampling moves due to the geometric constraints, the other is that as the system size grows the phase transitions become sharper such that for a hundred or more particles, thermal methods become essentially ineffective.

At high energies and moderate pressures entropy wins and all materials are gases, but as the energy decreases they typically condense into a liquid, then freeze into a solid, and sometimes even undergo solid-phase structural transitions.  
A specific heat curve is shown as an example in \cref{fig:PhaseDiagram} for the periodic Lennard-Jones system. 
In the thermodynamic limit, the discontinuity of the potential energy across the phase transition would result in a divergence of the heat capacity, but in a numerical simulation they are broadened by finite size effects and appear as sharp peaks. Performing the sampling at a range of pressure values, the loci of the peaks define the boundaries between stability regions of different phases in the pressure-temperature phase diagram. Notice the structural similarity of the solid phase, a closed packed face centred cubic (fcc) crystal, to the narrow global minimum of the \LJ{38} cluster. As the cluster size grows, the repeated occurrence of these fcc-like clusters hint at the ground state of the infinite crystal. There is no such periodic analogue for the broad icosahedral basin of the clusters because the five-fold symmetry is incompatible with periodicity (periodicity is further discussed in \cref{sec:periodic}).

NS has been successful in characterizing the behaviour of a wide range of materials. These include
model systems such as the Potts model\cite{potts,2018JCoPh.375..368P}, hard spheres~\cite{NS_hardsphere}, Lennard-Jones~\cite{pt_phase_dias_ns,Nielsen_npt_2015} and the Jagla potential~\cite{BartokJagla2021} as well as more chemically realistic potentials for aluminium and the shape memory alloy NiTi~\cite{ConPresNS}, as well as lithium~\cite{ns_lithium}, and iron~\cite{ns_iron}.  Recently, machine-learning interatomic potentials\cite{chemrevGPR} are now being applied~\cite{AgPd_ML} to increase the predictive power of calculated phase diagrams, leveraging the increased accuracy of the potentials. A detailed overview of materials applications can be found in \ccite{NS_materials_review}.

\section{Reproducibility and data deposition}\label{sec:repro}

We recommend a set of minimum considerations and reporting standards for NS computations, shown in \cref{box:checklist}. You should state clearly
the number of live points and stopping conditions. In addition, you should report implementation specific settings,
for example the number of repeats if using slice sampling, or the enlargement factor if using ellipsoidal sampling. If you are using a public software package,
report the version number. Since NS is an MC algorithm,
for reproducibility we suggest fixing the random seed so that identical results can be replicated.

Since NS computes an integral, make sure that the \gls{integrand} is explained clearly, including the choices of \gls{likelihood} and prior, and, for example, any overall constant factors that are sometimes omitted from the \gls{likelihood}. To help
achieve these goals, consider publishing your computer code alongside your research. Similarly, consider depositing the NS output files publicly. This would permit further scrutiny and re-use of the NS results. The output data should be accompanied with sufficient metadata (e.g., column labels; see \ccite{Wilkinson2016}).

Although the ultimate result might be for example, a ratio of NS results (a Bayes factor), we recommend reporting the results of all individual NS calculations. To do so,
we suggest the triplet of $\log\ev$, the estimated uncertainty, and the KL divergence, $H$. The first two are most relevant for inference, whereas the KL divergence indicates the numerical challenge, as it impacts runtime (\cref{eq:runtime}) and uncertainty (\cref{eq:delta_Z}). You may also wish to report the effective dimensionality or an Occam factor~\cite{Hergt:2021qlh}.

The NS error estimates are usually reliable, such that if we wish to reproduce NS computations, we should find agreement within uncertainties with the original calculation. As discussed previously, though, it is $\log\ev$ rather than $\ev$ alone that is distributed with a roughly symmetric Gaussian error.

\begin{boxedtextlhs}[box:checklist]{Check-list for reliability and reproducibility}
\begin{enumerate}[wide, labelwidth=0pt, labelindent=0pt]
    \item Pick an appropriate technique for sampling from the  \gls{constrained prior}; some choices perform more reliably and efficiently in high-dimensions
    \item Pick appropriate settings and tailor them to your goals and problem. Relaxed exploration settings (for example fewer steps\cite{polychord}) may be adequate if you are only concerned about parameter inference.
    \item Report software version numbers and settings
    \item Describe \glspl{prior} and \gls{likelihood} in adequate detail
    \item Ideally, publish the code used for the computation, permitting the calculation to be replicated and scrutinized
    \item Perform cross-checks on NS run~\cite{Fowlie:2020mzs}, as implemented in e.g., \code{anesthetic}. If practical, consider simulation-based calibration to check results
    \item Report triplet of $\log\ev$, the associated uncertainty, and $H$ for each NS computation
    \item Consider publishing the NS output, allowing further re-use and checks of the NS run
\end{enumerate}
\end{boxedtextlhs}

\section{Limitations and optimizations}\label{sec:limitations}
\subsection{Limitations}
Although NS is broadly applicable, there are several potential limitations. The first limitation relates to the prior: we sample from the  \gls{constrained prior} and thus require a proper prior.
For many NS implementations, this proper prior must be transformed from the unit hypercube. Normalizing flows were recently proposed~\cite{2021arXiv210212478A} for cases where this is inconvenient or impossible analytically, including using the \gls{posterior} from an NS run as a prior.

There are, furthermore, limitations related to the \gls{likelihood}. Integration using NS requires a non-negative \gls{integrand} (though the compression itself makes no such restriction). Whilst this condition is always fulfilled in statistical applications, it could be violated when NS is used as a general purpose integrator. We wrote \cref{eq:lebesgue} assuming that $\like \ge 0$ whereas \cref{eq:riemann} makes no restriction. NS, however, compresses upwards in \gls{likelihood}, such that positive and negative \gls{likelihood} regions of the integral would be treated differently, with the latter explored at inadequate resolution.
NS, furthermore, requires a \gls{tractable} \gls{likelihood}. For cases in which the \gls{likelihood} is intractable, \ccite{ns_systems_bio} proposed a \gls{likelihood}-free NS in the context of systems biology, assuming that an unbiased estimator of the \gls{likelihood} was available.

Lastly, plateaus in the \gls{likelihood} function spoil the estimates of the compression~\cite{Skilling:2006:nested,murray,riley,2020arXiv200508602S}. That is, sets $A$ with non-negligible prior mass, $\mu(A) > 0$, and constant \gls{likelihood}, which means $\like(\params) \equiv c$ for all $\params \in A$. We may modify the \gls{likelihood} function to remove plateaus by adding a negligible unique iid tie-breaking random draw to the \gls{likelihood} or promoting that draw to a parameter and increasing the dimension of the problem.  We may alternatively modify the algorithm itself such that it sums plateaus correctly but reduces to the ordinary NS algorithm in their absence. In the presence of plateaus, the minimum \gls{likelihood} among the live points, $\min \like$, may be shared by several live points.

For example, we could modify NS by removing all $q$ points in the plateau and then replacing them all~\cite{Fowlie:2020gfd}. In this case, the compression factors follow $\betadist(\nlive + 1 - q, q)$. This may be applied retrospectively to any NS runs, though suffers from increased uncertainty as the number of live points is temporarily reduced to as few as $\nlive - q$. Alternatively, we could first increase the live points by sampling subject to $\like > \threshold$ until $\nlive - 1$ lie at $\like > \min \like$. Note that $\min \like$ may decrease as we add live points at $\min \like > \like > \threshold$. We then remove $q$ points in the outermost contour. The compression factor follows $\betadist(\nlive, q)$. This cannot be applied retrospectively but the uncertainties are reduced (at the cost of greater run-time) as the number of live points is temporarily increased beyond $\nlive$. This may in fact be seen as a dynamic version of the first modification and we could increase the number of live points according to a different criterion. Lastly, we could split the integral into an integral over plateaus and an integral without plateaus, and perform only the latter with NS~\cite{2020arXiv200508602S}.

In addition, there are computational limitations. In ordinary NS running for longer doesn't reduce uncertainties or increase the number of \gls{posterior} samples.
If one is only interested in using NS for optimization, this isn't a drawback. In any case,  as discussed in \nameref{sec:dns}, dynamic NS overcomes this problem by rewinding and resuming an NS run. Besides that, NS requires \order{\dimension^2} iterations (\cref{eq:runtime}), the memory required to store the co-ordinates of every dead point scales as \order{\dimension^3} (with potentially nasty scaling factors for clustering in particular exploration strategies). This means that NS implementations that store every dead point become memory bound at \order{500} dimensions.

We stress, furthermore, that users should be aware of ways in which NS may fail:
\begin{enumerate}
    \item\label{item:fail} NS may fail to successfully draw independent samples from the  \gls{constrained prior}. Consequently NS results including error estimates may be faulty and anticipated properties of NS, such as convergence, won't hold. In \nameref{sec:results}, we discussed cross-checks on NS results, such as \ccite{Fowlie:2020mzs}, that may identify this issue.
    
    \item Owing to \cref{item:fail}, NS may miss \glspl{mode}. \gls{multi-modal} problems pose challenges in Bayesian computation, particularly in \gls{Markov chain Monte Carlo}, where the chain must make a sequence of unlikely steps between \glspl{mode}. Unfortunately, it is in general impossible to know whether all \glspl{mode} have been found. Broadly speaking, as NS does not depend on slow transitions between \glspl{mode}, it is well-suited to \gls{multi-modal} problems. Once a \gls{mode} is established, furthermore, it won't be abandoned until the \gls{likelihood} threshold exceeds that of the points that were in the mode.
    
    \item NS may sample inefficiently from the  \gls{constrained prior}. As discussed in \nameref{sec:exploration}, this may occur in high\hyph{}dimensional problems with rejection sampling strategies.
\end{enumerate}
Our check-list in \cref{box:checklist} includes checks of NS failures.

\subsection{Optimizations}
There are several ways NS runs may be optimized to make best use of computing resources (see also \nameref{sec:parallel}). First, one may utilize fast and slow parameters by breaking the \gls{likelihood} function into fast and slow operations that involve subsets of the $\dimension$ parameters. The parameters associated with fast and slow operations are referred to as fast and slow parameters, respectively; for example, if the \gls{likelihood} function may be written as,
\begin{equation}
    \like(x, y, z, \ldots) = \text{slow}(x) \times \text{fast}(y, z, \ldots),
\end{equation}
then $y, z \ldots$ are fast parameters and $x$ is a slow parameter.  When selecting a new point, to minimize runtime we should where possible change the fast parameters and keep the slow parameters constant, allowing caching of the slow operation~\cite{PhysRevD.87.103529}. This is natural in the slice sampling exploration strategy discussed in \nameref{sec:exploration}, as we may pick slices along fast directions~\cite{polychord}. If $\dimension_s$ parameters are slow, we require $\dimension_s$ rather than $\dimension$ slow operations per iteration. Similarly, it may be exploited in modified Metropolis algorithms~\cite{Au_2001, Lewis:2002ah} in which the proposals change blocks of parameters at a time.

In addition, one may repartition the \gls{posterior}. NS must compress exponentially through the entire prior volume, which could be slow for a diffuse prior with substantial compression to the \gls{posterior}. This in practice limits NS to \order{100s-1000s} of dimensions. This prohibits fitting complex hierarchical Bayesian models and deep neural networks, or reaching HMC like dimensionalities of \order{10^6}. In some cases, it may be possible to use a narrower prior and correct the \gls{evidence} estimates from NS post-hoc.

This issue may occur in problems in which the prior is unrepresentative~\cite{chen2019improving}, i.e., where the observed data lies in the tail of the prior predictive distribution. This may be mitigated by Bayesian automatic prior repartitioning~\cite{chen2019bayesian} in which one redefines the prior and \gls{likelihood} while leaving their product unchanged. This allows one to keep the \gls{evidence} integral in \cref{eq:Z} unchanged, but reduce the prior to \gls{posterior} compression in \cref{eq:kl}.
For example, consider a Gaussian \gls{likelihood}, $\normaldist(x, \sigma^2)$, for a parameter $x \sim \uniformdist(-L / 2, L / 2)$ for $\sigma \lll L$. The compression would be approximately $H \approx \log(\sigma / L)$. Repartitioning the prior and \gls{likelihood} by swapping them, we obtain $H \approx 0$.

The possibility of improving the robustness and efficiency of NS by exploiting the intrinsic degeneracy between the `effective' likelihood and prior in the formulation of Bayesian inference problems was also discussed in~\ccite{Feroz_2009_SZ}, and posterior repartitioning can further be viewed as the vanilla case (when the importance weight function equals to $1$) of nested importance sampling proposed in~\ccite{Chopin_2010}.

Lastly, rather than performing new NS runs, one may reuse runs for similar \glspl{likelihood} and \glspl{prior}. If the prior or \gls{likelihood} are modified, it may be possible to reweight the \gls{posterior} weights and \gls{evidence} integral~\cite{10.1214/13-STS465},
\begin{align}
    \post_i^\prime &= \post_i \times \frac{\like^\prime_i \pi^\prime_i}{\ev^\prime}\frac{\ev}{\like_i \pi_i}  \\
    \ev^\prime &= \ev \times \sum \post_i \times \frac{\like^\prime_i \pi^\prime_i}{\like_i \pi_i}
\end{align}
where $\post_i = w_i \like_i$. This re-weighting may be interpreted as a \gls{pseudo-importance sampling} in which the original estimated \gls{posterior} plays the role of the kernel. This is particularly useful for investigating prior sensitivity in the context of Bayesian inference. The effective sample size of draws from the new \gls{posterior} may be used to judged the reliability of this procedure; if the new and original \gls{posterior} distributions differ substantially, a fresh NS run needs to be performed.

There may, furthermore, be cases in which we wish to investigate several similar \gls{likelihood} functions at once; \ccite{buchner2019bigdata} presents a collaborative version of NS that operates on more than one \gls{likelihood} function at once and in which parts of the \gls{likelihood} evaluation may be recycled.

\section{Outlook}\label{sec:outlook}

The \gls{Markov chain Monte Carlo} computational revolution of the 1990s solved the problem of computing shapes. Skilling's remarkable NS algorithm solved the outstanding problem of computing magnitudes at the same time. Although it is naturally expressed in the language of Bayesian inference, NS is a powerful and general-purpose integration algorithm. For that reason, we expect NS to remain relevant long into the future. We are now in the midst of a revolution in data science, and high-dimensional spaces and integration are more important than ever. NS rises to the challenge.

As we reviewed, there have been many theoretical developments in understanding NS, including its convergence, errors, diagnostics and techniques for sampling from the  \gls{constrained prior}. We expect them to continue, especially as connections to other statistical methods and machine learning are explored. Indeed, theoretical analysis of NS combined with constrained \gls{Markov chain Monte Carlo} exploration may be helped by the connections to SMC discussed in \cref{box:smc}, and we anticipate further developments in understanding uncertainties in NS, especially for parameter inference and in the case of a dynamic number of live points. Lastly, NS may be considered a meta-algorithm, as it doesn't specify an algorithm for sampling from the  \gls{constrained prior}. We are already seeing that this opening allows NS to dovetail with developments in machine learning, such as normalising flows that are beginning to be used to sample from the  \gls{constrained prior}.

The successes and breadth of applications of NS stem from the fact that it is fundamentally simple. Whilst in the future NS may be understood and improved in the context of more sophisticated computational methods, NS will retain advantage and appeal due to its simplicity.
\section*{Acknowledgements}
\setlength{\parindent}{0pt}
\emph{We thank John for his wonderful algorithm. The success of nested sampling may be that simple beats clever; but the beauty of nested sampling is that it is both simple and clever.}
\parskip = \baselineskip

We thank Kyle Barbary for discussions. AF was supported by an NSFC Research Fund for International Young Scientists grant 11950410509.
LBP acknowledges support from the EPSRC through an Early Career Fellowship (EP/T000163/1). MH acknowledges support from the Carl Zeiss Foundation. NB was funded by the U.~S.\ Naval Research Laboratory's base 6.1 research program, and CPU time from the U.~S.\ DoD's HPCMPO at the AFRL and ARL DSRCs. MP acknowledges support from the STFC (ST/V001213/1 and ST/V005707/1).
WH was supported by a Royal Society University Research Fellowship. 

\section*{Author Contributions}
Introduction (A.~F., D.~S., L.~S., P.~W.); Experimentation (J.~B., E.~H., J.~S.~S.); Results (A.~F.); Applications (G.~A., N.~B., X.~C., G.~C., F.~F., M.~G., W.~H., M.~H., A.~L., D.~P., L.~B.~P., M.~P., J.~V., D.~W., D.~Y.); Reproducibility and data deposition (A.~F.); Limitations and optimizations (A.~F.); Outlook (A.~F.); Overview of the Primer (A.~F.). 

\section*{Competing Interests}
There are no competing interests to declare.

\appendix

\hypertarget{target:supp}{\section*{Supplementary Information\label{sec:supp}}}

\printglossaries

\section{Estimators for compression factor}\label{sec:compression}

Besides the estimator $\langle \log \compression \rangle  = -\niter / \nlive$ in \cref{eq:geometric}, we may instead consider,
\begin{equation}
\langle \compression \rangle = \frac{\nlive}{\nlive + 1}.
\end{equation}
Similarly, \ccite{walter2015point} suggests the estimator
\begin{equation}
\hat\compression = 1 - \frac{1}{\nlive}.
\end{equation}
In any case, the relative differences in $\log \compression$ are of order $\order{1/\nlive}$.

\section{Historical background}\label{app:annealing}

In the 1950s, when computational algorithms were beginning to be developed, the Bayesian revolution lay decades in the future and inquiry was focused at best on \gls{posterior} distributions. The first practical computer codes for this were \gls{Markov chain Monte Carlo}~\cite{2008arXiv0808.2902R}; first, the Metropolis algorithm~\cite{Metropolis:1953} and later Metropolis-Hastings\cite{Hastings:1970}. They build a \gls{Markov chain} of correlated points drawn from an unnormalized \gls{posterior}. The Metropolis algorithm accepted a transition
$\params\rightarrow\params'$ only if
\begin{equation}\label{eq:mcmc}
\like(\params') > u\like(\params)
\end{equation}
 where $u\sim \uniformdist(0,1)$.
This conditional acceptance of random trial transitions satisfies detailed balance and converges towards an equilibrium.
This, however, only gives the shape of the distribution, not the normalizing constant, $\ev$.

\begin{figure*}[thb!]
\centering
  \includegraphics[width=0.9\textwidth]{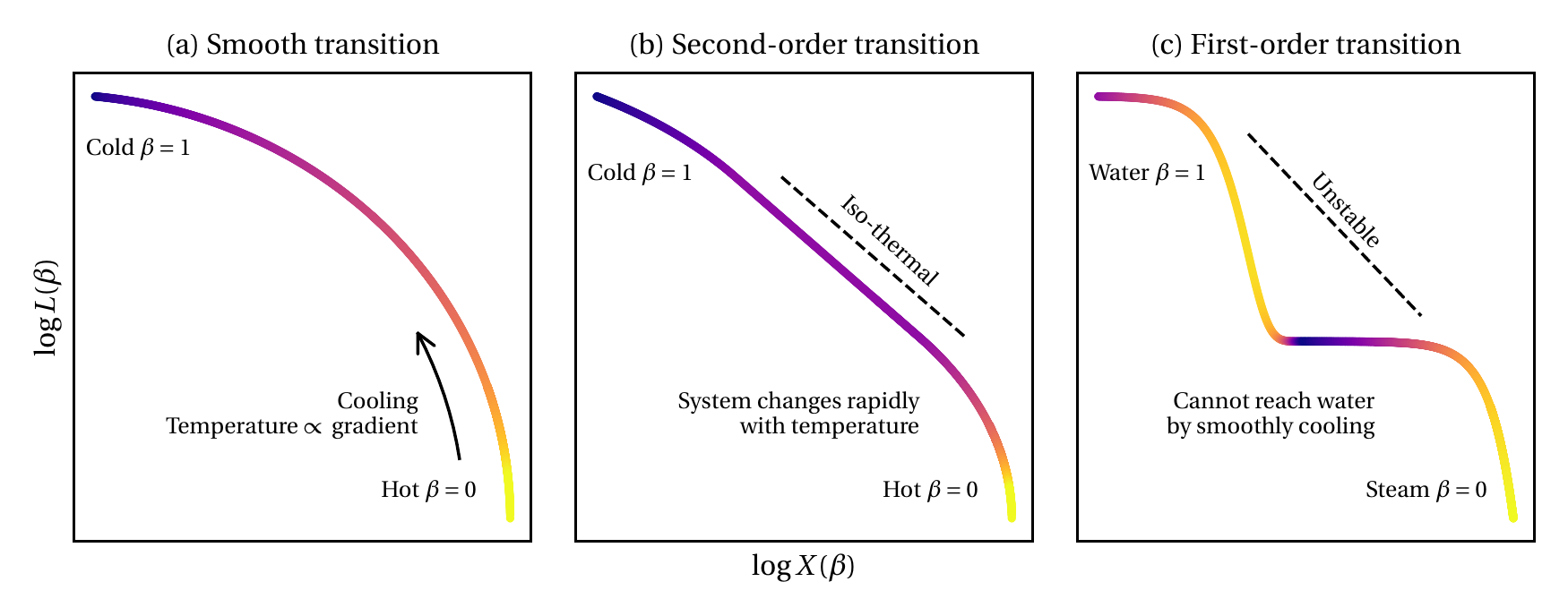}
\caption{\captiontitle{Phase transitions in simulated annealing} We show $\log\like(\beta)$ and $\log X(\beta)$ as we cool from $\beta = 0$ to $1$ in an ordinary application, a second-order transition and a first-order transition. The colour scale indicates temperature.}
\label{fig:PhaseChanges}
\end{figure*}

Later, in the 1970s, a connection between the \gls{likelihood} and the temperature-dependent energy,
\begin{equation}
\like(\params) \propto e^{- \beta E(\params)}
\end{equation}
where $\beta = 1/ T$, suggested generalizing the \gls{likelihood} to $\like^\beta$. Upon which, the \gls{evidence} generalized to
\begin{equation}\label{eq:generalised-evidence}
\ev(\beta) = \int \like^\beta(\params) \prior(\params) \, \diff\params,
\end{equation}
with $\ev = \ev(1)$ being the desired \gls{evidence} and $\ev(0)$ being 1. Whereas the \gls{posterior} generalized to
\begin{equation}
\post^\beta(\params) = \frac{\like^\beta(\params) \prior(\params)}{\ev(\beta)},
\end{equation}
with $\post^0(\params)$ the prior and $\post^1(\params)$ the desired \gls{posterior}.
These values are connected by the computable differential
\begin{equation}
  \frac{\diff \log \ev}{\diff \beta} = \int \log \like(\params)  \post^\beta(\params) \,\diff\params = \big\langle \log \like \big\rangle_\beta
\end{equation}
which is simply the log-likelihood averaged over the distribution $\post^\beta(\params)$ at inverse temperature $\beta$.
Add that up as the distribution is slowly cooled from $\beta = 0$ to $1$, i.e., from the prior to the \gls{posterior}, and the \gls{evidence} arrives as~\cite{doi:10.1063/1.1749657,Kirkpatrick+etal:1983,10.1214/ss/1028905934}
\begin{equation}
  \log \ev = \int_0^1 \left\langle \log\like \right\rangle_\beta\,\diff\beta.
\end{equation}
All that seems to be needed to compute the \gls{evidence} integral is to fix the cooling schedule to track $\beta$ suitably slowly (hence annealing) from 0 to 1 such that the \gls{posterior} distribution $\post^\beta(\params)$, represented by a collection of samples drawn from it, changes slowly. This annealing schedule may, however, be problematic. Maintaining the thermodynamics analogy, which is further expanded in \cref{box:stat_mech}, the trouble occurs with changes of state as the system cools. The \gls{evidence} plays the role of the thermodynamic \gls{partition function} and at each inverse temperature the samples are expected to coalesce closely around the local maximum of the \gls{posterior} distribution $\post^\beta(\log X)$. This implies that
\begin{equation}\label{eq:coalesce}
\frac{\diff \log \like(X)}{\diff \log X} = -\frac{1}{\beta}
\end{equation}
In smooth cross-over transitions, volume and energy shrink in tandem as the system cools from $\beta = 0$ (vertical slope) to $\beta = 1$ (diagonal slope) in \cref{fig:PhaseChanges}-a, which is what happens in stable physical systems.

If there is a second-order phase transition, however, a tiny cooling around the critical $\beta$ rapidly compresses the ensemble exponentially all the way from disorder to order. We see this in the iso-thermal region in \cref{fig:PhaseChanges}-b.
The sensitivity of the system to temperature means that it becomes numerically impossible to use $\beta$ as the control parameter. Lastly, if there's a first-order phase transition, two phases coexist at the same temperature or in a temperature interval, e.g., steam and water in \cref{fig:PhaseChanges}c. Starting from the steam phase, to reach the water phase requires an abrupt first-order phase transition. It thus becomes impossible even in principle to use inverse temperature to change smoothly from the steam phase to the water phase.

The first-order phase transition was connected to the fact that there were multiple solutions to \cref{eq:coalesce} for some $\beta$. This implies that the \gls{integrand} in the \gls{evidence} integral expressed as
\begin{equation}
\ev(\beta) = \int \like^\beta(X) X \, \diff \log X,
\end{equation}
was \gls{multi-modal} for at least some temperatures. Physically, this may occur in a problem with distinct solutions at different \gls{likelihood} levels. In our analogy, we may interpret the \glspl{mode} as phases. Traditional annealing methods struggle to move particles between them during a phase transition but with NS, the system's evolution can be controlled through compression instead.

\section{Step sampling}\label{sec:step_sampling}

For detailed balance, the transition kernel in a step sampler must satisfy
\begin{equation}\label{eq:detailed_balance}
\int P(\params_f | \params) \, \prior^\star(\params) \, \diff \params = \prior^\star(\params_f).
\end{equation}
We are thus drawing a new live point $\params_f$ from
\begin{equation}\label{eq:walk}
P(\params_f | \left\{\params_\text{live}\right\}) = \frac{1}{\nlive} \sum_{i=1}^{\nlive} P(\params_f | \params^i_\text{live})
\end{equation}
where $\left\{\params_\text{live}\right\}$ denotes the set of live points and we average over the possible choices of initial live point indexed by $i$, $\params_\text{live}^i$.
For a draw from \cref{eq:walk} to be approximately equivalent to an independent draw from the  \gls{constrained prior}, we thus require
\begin{equation}\label{eq:walk_approx}
P(\params_f | \left\{\params_\text{live}\right\}) \approx \prior^\star(\params_f).
\end{equation}
We anticipate that this might approximately hold because the live points should be independent draws from the  \gls{constrained prior} such that \cref{eq:walk} is a Monte Carlo estimate of the  \gls{constrained prior} through \cref{eq:detailed_balance}. In any case, to satisfy \cref{eq:walk_approx}, we may attempt to satisfy a condition of decorrelation between the chosen initial point and the final point,
\begin{equation}
P\left(\params_f | \params^i_\text{live}\right) \approx \prior^\star(\params_f).
\end{equation}
In \gls{uni-modal} problems, this can be achieved by performing a sufficient number of steps. In \gls{multi-modal} problems, it may be impossible as walks between \glspl{mode} are all but impossible. Since we only require \cref{eq:walk_approx}, however, correlation between the \glspl{mode} in which the initial and final points lie may be acceptable. The number of steps required to satisfy \cref{eq:walk_approx} may decrease when $\nlive$ is increased, as Monte Carlo errors in the approximation in \cref{eq:walk_approx} shrink as $1/\sqrt{n}$.

Lastly, these considerations show that starting from the last dead point instead of a randomly chosen live point isn't sensible~\cite{potts}, even though it avoids correlation with the existing live points. In \gls{uni-modal} problems, we could require many steps to avoid biasing the draw towards the lowest likelihood levels near the last dead point. In \gls{multi-modal} problems, it would lead to catastrophe, as the fractions of live points in each mode could not change, since a point would always be replaced by a point in the same mode. Consequently, modes would never die and NS would become stuck at the likelihood threshold of the lowest lying mode.

\section{Uncertainty in the posterior}\label{sec:uncertainty_posterior}

There is a further source of uncertainty in \gls{posterior} expectations. From the dead points $\params_i$, with importance
weights $\post_i$, a \gls{posterior} mean for some quantity of interest $f:\Omega\to \mathbb R$ may be estimated by~\cite{Chopin_2010}
\begin{align}
\langle f(\params) \rangle &= \int
f(\params) \post(\params) \, \diff\params = \int f(X) \post(X) \, \diff X\\
& \approx \sum_i \post_i f(X_i) \approx \sum_i \post_i f(\params_i),\label{eq:posterior_mean}
\end{align}
where $f(X)$ is the mean of $f$ over an \gls{iso-likelihood contour},
\begin{equation}
f(X)
= \frac{\int f(\params) \delta(\like(\theta) - \like(X)) \prior(\params) \, \diff \params}{\int \delta(\like(\theta) - \like(X)) \prior(\params) \, \diff \params}.
\end{equation}
There are thus two sources of uncertainty in \cref{eq:posterior_mean}: the statistical uncertainty in the estimates of the importance weights $\post_i$, as before, and the fact that \cref{eq:posterior_mean} replaces the mean around the contour by a single draw from around the contour, $f(X_i) \approx f(\theta_i)$. As before, the first source can be accounted for by simulating
the compression factors through \cref{eq:beta}. Whilst both sources are reduced by increasing the number of live points, the second source usually dominates.

\Ccite{Higson2018sampling} demonstrated that it can be assessed by decomposing an NS run with $\nlive$ live points into $\nlive$ NS runs
with one live point each called \emph{threads}. The threads can be recombined in
many different combinations using resampling methods such as \gls{bootstrap} to generate
simulated runs with the same number of live points as the
original run. 
The variance in \gls{posterior} inferences in these simulated runs captures both sources of uncertainty (see \code{UltraNest}\cite{2021JOSS....6.3001B} for an NS implementation of these simulations).

\section{Periodic boundary conditions, degrees of freedom and Monte Carlo moves}\label{sec:periodic}

In any condensed phase, a computationally tractable small finite system would be
dominated by surface effects, and eliminating these requires a periodic
supercell description.  The typical volume per atom changes by orders of magnitude between
the gas and condensed phases, and must be allowed to vary for NS to sample the relevant structures.
This can be achieved by fixing the pressure $P$, replacing the potential energy $U$ in the sampling algorithm with the
enthalpy $U + PV$, where $V$ is the system volume, and sampling the cell degrees of freedom~\cite{pt_phase_dias_ns,ConPresNS}.
The corresponding \gls{partition function} is
\begin{align}
 Z(N, P, \beta) & =  Z_p \beta P \int d\mathbf{h}_0\, \delta(\det \mathbf{h}_0 - 1) \times \nonumber \\
 & \int_0^\infty dV V^N \int_{(0,1)^{3N}} d\mathbf{s}\, e^{-\beta (U(V^{1/3} \mathbf{h}_0 \mathbf{s}) + P V)}, \nonumber
\end{align}
where $\mathbf{h}_0$ is the reduced cell shape matrix (with unit determinant) and $\mathbf{s}$ are the scaled (often called fractional) atomic positions. When generating uniformly
distributed configurations $V$ and $\mathbf{h}_0$ must be sampled in addition to the scaled positions $\mathbf{s}$.
The expectation value of $V(\beta)$ must be calculated using the same weights as the \gls{partition function}.  To compute an entire
pressure-temperature phase diagram the NS run is repeated for each pressure value. After a completed NS run, the expectation value
of any observable can be calculated as a function of temperature, e.g.\ the radial distribution function plotted in \cref{fig:PhaseDiagram}, which can be compared to the results of scattering experiments and used to identify each equilibrium phase.

Condensed phase atomic position Monte Carlo moves with a reasonable acceptance probability are also challenging
to generate.  Single atom moves become inefficient if the resulting energy change cannot be computed in $\order{1}$
time, and the probability of accepting naive collective moves decreases as $1/\sqrt{N}$.
A more efficient alternative is Galilean Monte Carlo~\cite{Betancourt2011,Skilling2012,ConPresNS}, where a random direction in $3N$-dimensional configuration
space is proposed, the entire system is propagated for a fixed number of steps along straight lines that reflect specularly
from the allowed $U(q) < U^\star$ boundary.
Another is total-energy Hamiltonian Monte Carlo~\cite{Duane1987,ConPresNS}, where the kinetic energy is added to $U$, so the NS
iteration constrains \emph{total} energy, and atomic moves are proposed by carrying out short time constant energy molecular dynamics
trajectories that are accepted or rejected in their entirety. In both cases, Monte Carlo moves that propose to change the volume and the cell shape are also used. 

In condensed phase systems with multiple types of particles the probability for particles to switch places is low, especially in the solid phase, and
explicit particle swap proposals are required to ensure mixing.  Further, the \emph{composition}
of different phases may change discontinuously across phase transitions.
The semi-grand-canonical ensemble, with fixed total number of particles but variable composition, describes
this situation.  Monte Carlo proposals include changes to particle type, and the energy is augmented by $\sum_i \mu_i N_i$,
where $i$ runs over the types, $\mu_i$ is a specified chemical potential, and $N_i$ is the number of particles of that type $i$.  Like
$V(\beta)$ above,  $N_i(\beta)$ is an output of the simulation.

\section{Example}\label{sec:example}

\definecolor{forestgreen(web)}{rgb}{0.13, 0.55, 0.13}
\definecolor{fireenginered}{rgb}{0.81, 0.09, 0.13}
\definecolor{indigo(web)}{rgb}{0.29, 0.0, 0.51}
\lstset{%
  keepspaces,
  framerule=1pt,
  backgroundcolor=\color{white},
  basicstyle=\sffamily,
  breakatwhitespace=false,
  breaklines=true,                 
  frame=single,
  keywordstyle=\color{forestgreen(web)},
  language=Python,
  numbers=none,
  rulecolor=\color{boxedtext}, 
  showspaces=false,  
  showstringspaces=false,
  tabsize=2,
  columns=fullflexible,
  emph={},
  emphstyle=\color{fireenginered},
  morestring=*[b]',
  belowskip=-7pt,
  escapeinside={(*}{*)},
  commentstyle={\rmfamily\color{indigo(web)}}
}

\newcommand{\bash}{\color{boxedtext}{\$}}
\newcommand{\py}{\color{boxedtext}{$\ggg$}}
\newcommand{\nopy}{\color{boxedtext}{$\ldots$}}

Let us demonstrate NS using standard NS software on a toy problem using \code{Python}. The libraries used in this example may be installed using \code{pip}, for example,
\begin{lstlisting}[language=bash]
(*\bash*) pip install numpy scipy matplotlib pypolychord anesthetic 
\end{lstlisting}
To record our software version numbers, 
\begin{lstlisting}
(*\py*) import platform 
(*\py*) print(platform.python_version())
3.8.10
(*\py*) from importlib.metadata import version 
(*\py*) print(version('pypolychord'))
1.20.1
(*\py*) print(version('anesthetic'))
1.3.6
\end{lstlisting}
Let us attempt to compute the two-dimensional Gaussian integral,
\begin{equation}
    \ev = \frac{1}{(2 a)^2}\int_{-a}^a \int_{-a}^a e^{-x^2 - y^2} \,\diff x \diff y.
\end{equation}
We may write the \gls{integrand} in the form \cref{eq:Z} by defining the \gls{likelihood}
\begin{equation}\label{eq:like_example}
    \like(x, y) = e^{-x^2 - y^2}
\end{equation}
and the \gls{prior},
\begin{equation}\label{eq:prior_example}
    \prior(x, y) = \frac{1}{(2 a)^2}
\end{equation}
for $-a < x < a$ and $-a < y < a$ and vanishing elsewhere. So long as $a$ is greater than about $1$, $\ev \simeq \pi / (2 a^2)$ and $H \simeq -1 - \log\pi - \log 2a^2$. For concreteness, we take $a = 5$ such that $\log \ev \simeq -3.46$ and $H \simeq 2.46$ 

To run NS on this problem we first implement the logarithm of the \gls{likelihood} in \cref{eq:like_example},
\begin{lstlisting}
(*\py*) def loglike(theta):
(*\nopy*)      return -(theta**2).sum(), []  # [] is anything else to be saved
\end{lstlisting}
and a transformation of the unit hypercube representing the prior in \cref{eq:prior_example},
\begin{lstlisting}
(*\py*) def prior(unit_hypercube):
(*\nopy*)      a = 5. 
(*\nopy*)      return 2. * a * unit_hypercube - a
\end{lstlisting}
Here we map from the unit hypercube $\uniformdist(0, 1)^2$ to our $\uniformdist(-a, a)^2$ prior.

We use the \PC\cite{polychordcosmo,polychord} NS implementation --- this uses slice sampling to draw replacement live points from the constrained prior --- see \cref{tab:codes} for alternative NS software. We import it by
\begin{lstlisting}
(*\py*) from pypolychord import run_polychord
(*\py*) from pypolychord.settings import PolyChordSettings
\end{lstlisting}
We specify that our integral is two-dimensional (\code{ndim = 2}), that there are no derived quantities to save to disk (\code{nderived = 0}), and our NS settings. We wish to use $\nlive = 1000$ (\code{settings.nlive = 1000}), perform $5$ slice sampling steps (\code{settings.num\_repeats = 5}), and fix our random seed for reproducibility (\code{settings.seed = 67}).
\begin{lstlisting}
(*\py*) ndim = 2
(*\py*) nderived = 0
(*\py*) settings = PolyChordSettings(ndim, nderived)
(*\py*) settings.nlive = 1000  # this is (*\color{indigo(web)}{\nlive}*)
(*\py*) settings.num_repeats = 5
(*\py*) settings.seed = 67
\end{lstlisting}
Finally, we run the NS algorithm,
\begin{lstlisting}[mathescape]
(*\py*) run_polychord(loglike, ndim, nderived, settings, prior)
\end{lstlisting}
This by default writes our results to files named \code{chains/test*} and information to the screen. We may further inspect the results using, for example, \code{anesthetic}\cite{Handley:2019mfs}. First, we load the data from disk and for reproducibility fix the random seed using \code{numpy}\cite{harris2020array},
\begin{lstlisting}
(*\py*) from anesthetic import NestedSamples
(*\py*) import numpy as np

(*\py*) samples = NestedSamples(root='chains/test')
(*\py*) np.random.seed(71)
\end{lstlisting}
We may compute and print the triplet $\log\ev$, its uncertainty, and the KL divergence,
\begin{lstlisting}[mathescape]
(*\py*) H = samples.D()
(*\py*) logZ = samples.logZ()
(*\py*) uncertainty = (H / settings.nlive)**0.5  # this is (*\color{indigo(web)}{\cref{eq:delta_Z}}*)
(*\py*) print(logZ, uncertainty)
-3.5109005855438395 0.04994705730822035 
(*\py*) print(H)
2.494708533750648
\end{lstlisting}
Thus in this problem NS estimates $\log \ev = -3.51 \pm 0.05$ and $H = 2.5$ in agreement with the analytic results. We may compare these estimates to those found from $1000$ simulations (\code{nsamples=1000}) of the compression factors,
\begin{lstlisting}[mathescape]
(*\py*) draws = samples.logZ(nsamples=1000)
(*\py*) mean = np.mean(draws)
(*\py*) std = np.std(draws)
(*\py*) print(mean, std)
-3.510624723981491 0.05232090274229939
\end{lstlisting}
showing that the standard NS estimators are reliable in this case. We may check the assumption that $\log\ev$ is approximately Gaussian distributed by histogramming the draws of $\log\ev$. Here we use the \code{matplotlib}\cite{Hunter:2007} histogramming functionality and the \code{scipy}\cite{2020SciPy-NMeth} implementation of the normal distribution,
\begin{mdframed}[linewidth=1pt,linecolor=boxedtext,innerleftmargin=2pt,leftmargin=-4pt,innertopmargin=10pt]%
\begin{lstlisting}[mathescape,frame={}]
(*\py*) import matplotlib.pyplot as plt
(*\py*) from scipy.stats import norm

(*\py*) plt.hist(draws, bins='auto', density=True)
(*\py*) x = np.linspace(mean - 5. * std, mean + 5. * std, 1000)
(*\py*) plt.plot(x, norm(mean, std).pdf(x))
(*\py*) plt.show()
\end{lstlisting}%
\includegraphics[width=0.75\textwidth]{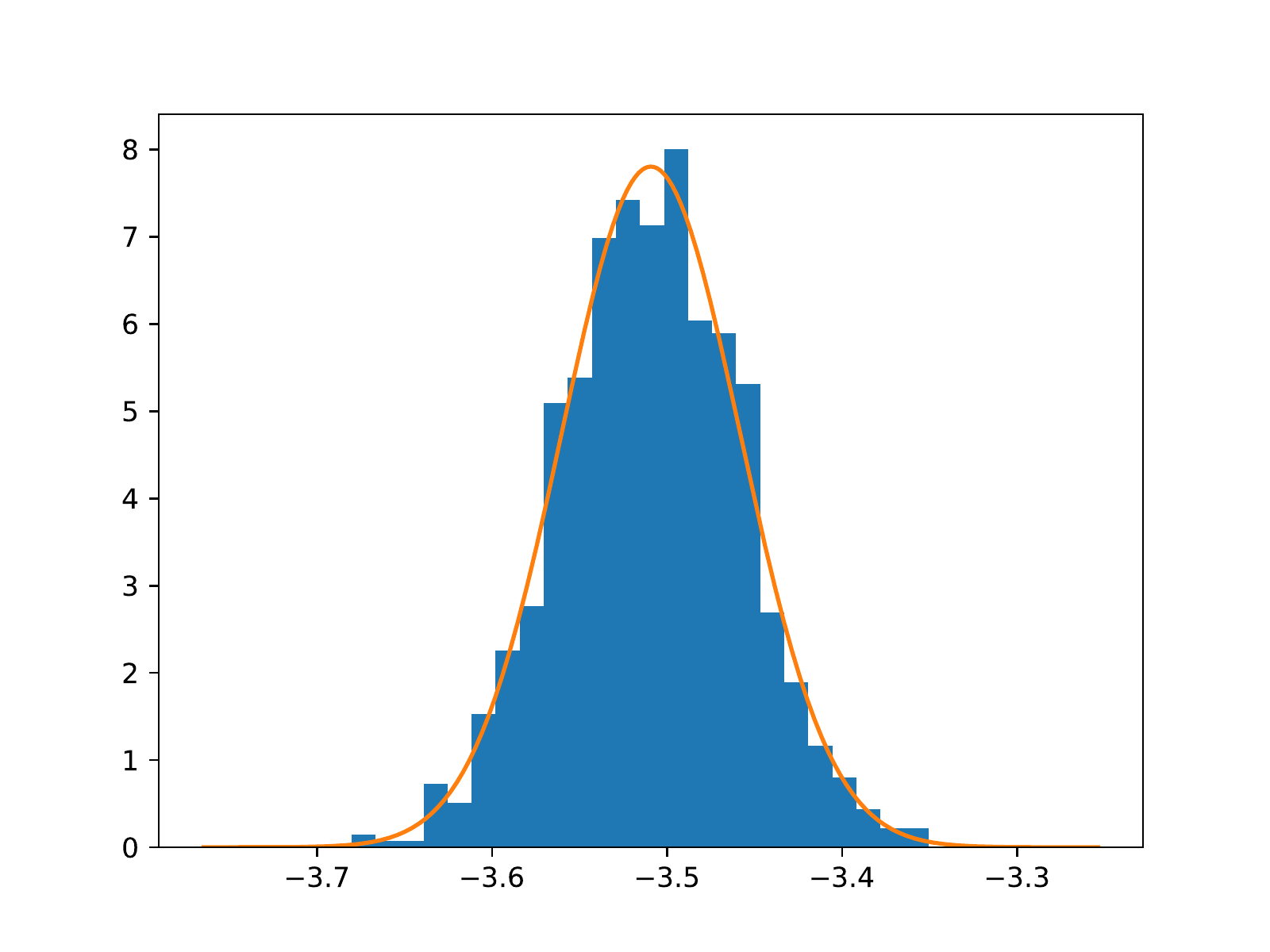}%
\end{mdframed}
We see that $\log \ev$ approximately follows a Gaussian distribution.

As a cross-check, we can test whether the insertion indexes of new live points are uniformly distributed using \code{anesthetic},
\begin{lstlisting}[mathescape]
(*\py*) from anesthetic.utils import insertion_p_value
(*\py*) samples._compute_insertion_indexes()
(*\py*) ks = insertion_p_value(samples.insertion, settings.nlive)
(*\py*) print(ks['p-value'])
0.4965573665241407
\end{lstlisting}
In this case we find $p$-value of about $0.5$, which does not indicate any discrepancy from uniform.

Lastly, we may wish to compute and plot posterior distributions. Here we plot one- and two-dimensional posterior distributions using \code{anesthetic} and kernel density estimation (\code{'kde'}),
\begin{mdframed}[linewidth=1pt,linecolor=boxedtext,innerleftmargin=2pt,leftmargin=-4pt,innertopmargin=10pt]%
\begin{lstlisting}[mathescape,frame={}]
(*\py*) samples.plot_2d(samples.columns[:ndim], types={'lower': 'kde', 'diagonal': 'kde'})
(*\py*) plt.show()
\end{lstlisting}%
\includegraphics[width=0.75\textwidth]{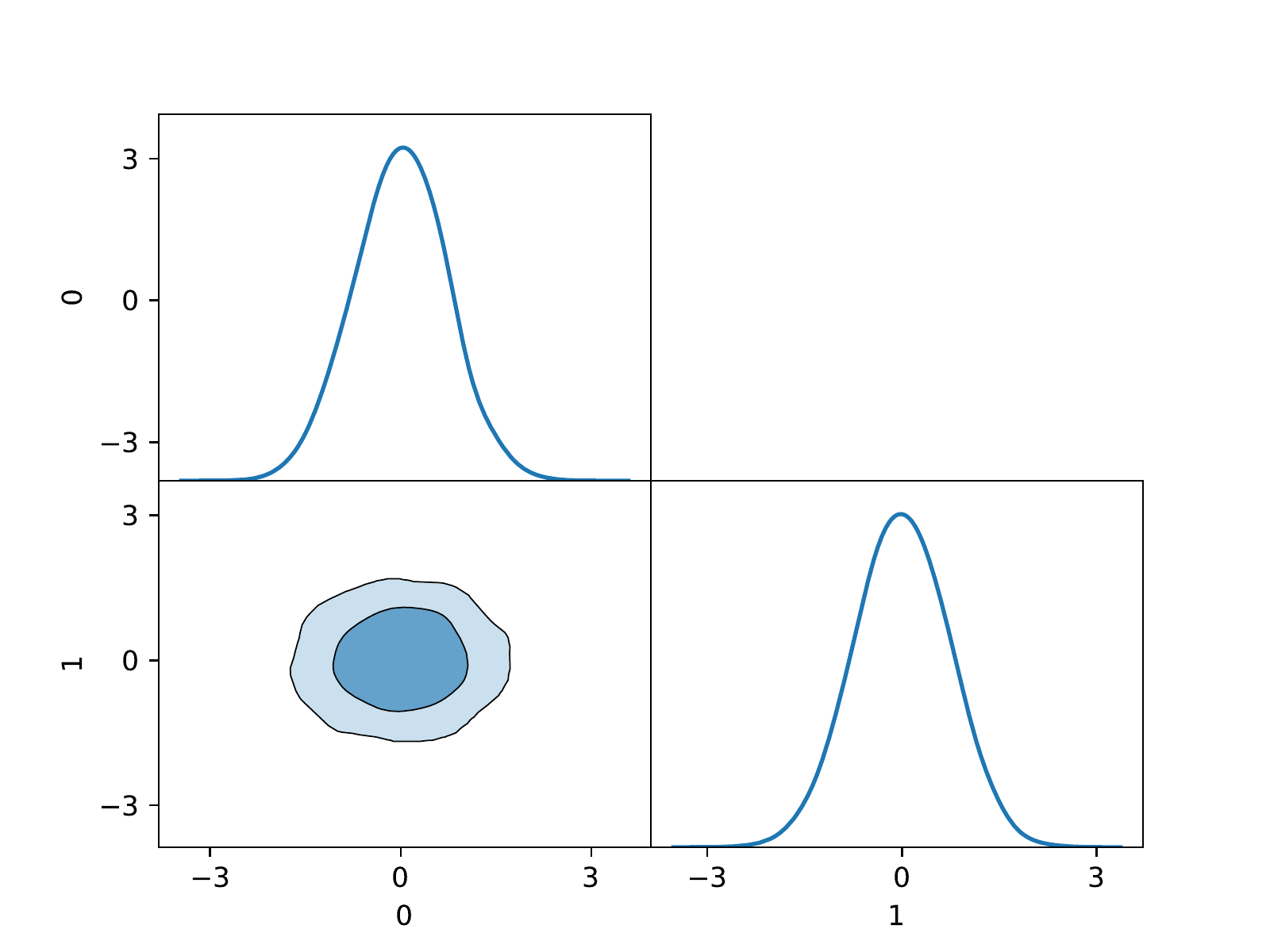}%
\end{mdframed}
We may examine the evolution of live points interactively using a graphical user interface (GUI)
\begin{mdframed}[linewidth=1pt,linecolor=boxedtext,innerleftmargin=2pt,leftmargin=-4pt,innertopmargin=10pt]%
\begin{lstlisting}[mathescape,frame={}]
(*\py*) gui = samples.gui()
(*\py*) gui.param_choice.buttons.set_active(1)  # show both parameters
(*\py*) gui.evolution.slider.set_val(3000)  # shows run after 3000 iterations
(*\py*) plt.show()
\end{lstlisting}%
\includegraphics[width=0.75\textwidth]{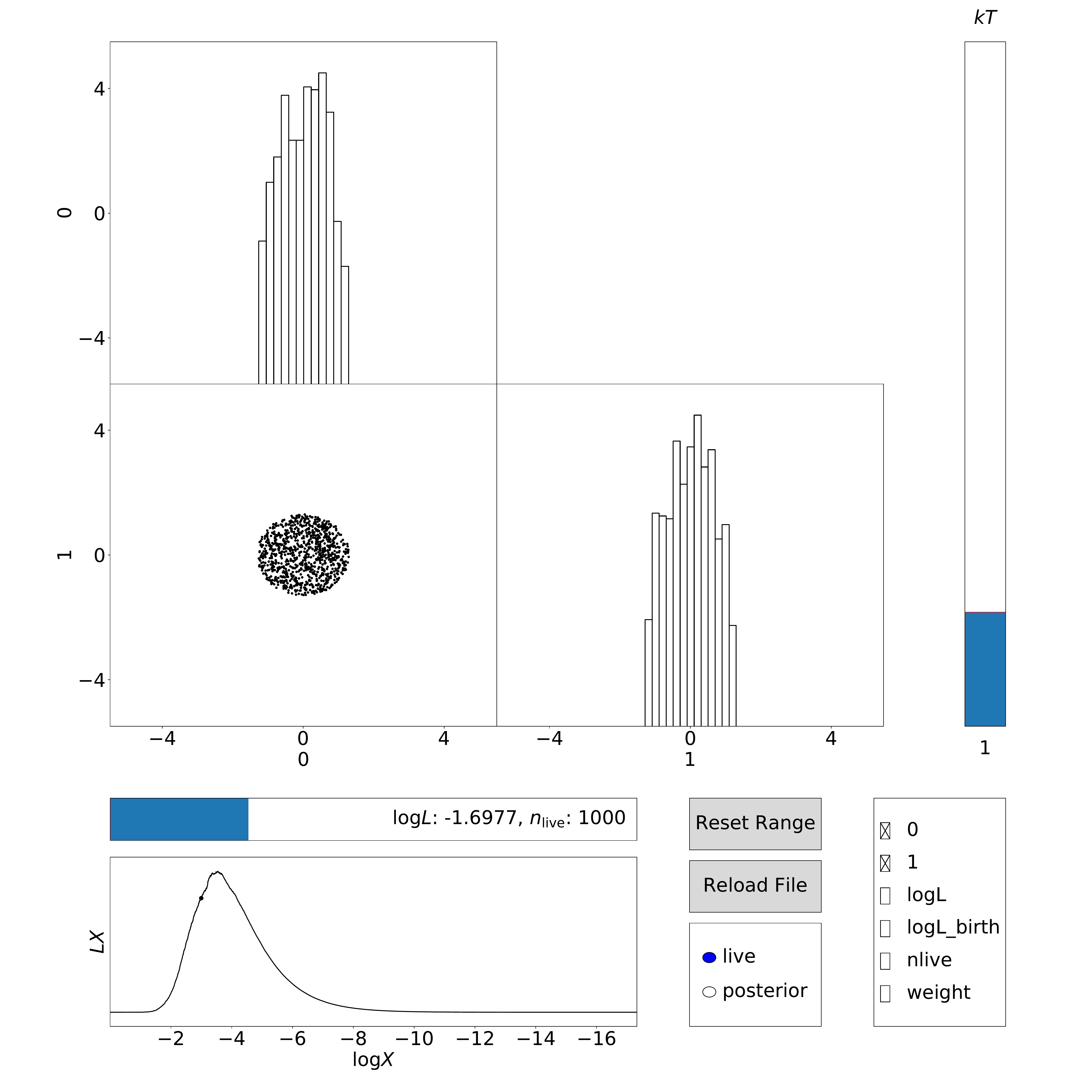}%
\end{mdframed}
This shows, among other things, the distribution of live points at iteration $3000$, and we may use the slider to see how it evolved.

\bibliography{references}
\bibliographytable{references}
\addcontentsline{toc}{chapter}{References}

\end{document}